\documentclass{SciPost}

\hypersetup{
    colorlinks,
    linkcolor={red!50!black},
    citecolor={blue!50!black},
    urlcolor={blue!80!black}
}

\usepackage[bitstream-charter]{mathdesign}
\usepackage{graphicx}% Include figure files
\usepackage{dcolumn}% Align table columns on decimal point
\usepackage{bm}% bold math
\usepackage{comment}
\usepackage{hyperref}
\usepackage{xcolor}
\usepackage{slashed}
\usepackage[normalem]{ulem}

\newcommand \bea{\begin{eqnarray}}
\newcommand \eea{\end{eqnarray}}

\newcommand\rvec{{\bf r}}

\def\simge{\mathrel{%
       \rlap{\raise 0.511ex \hbox{$>$}}{\lower 0.511ex \hbox{$\sim$}}}}
\def\simle{\mathrel{
       \rlap{\raise 0.511ex \hbox{$<$}}{\lower 0.511ex \hbox{$\sim$}}}}
\def\beq{\begin{equation}}
\def\eeq{\end{equation}}

\newcommand{\rv}{\mathbf{r}}

\definecolor{green}{rgb}{0.0,0.5, 0.0}

\DeclareSymbolFont{usualmathcal}{OMS}{cmsy}{m}{n}
\DeclareSymbolFontAlphabet{\mathcal}{usualmathcal}

\fancypagestyle{SPstyle}{
\fancyhf{}
\lhead{\colorbox{scipostblue}{\bf \color{white} ~SciPost Physics }}
\rhead{{\bf \color{scipostdeepblue} ~Submission }}

\fancyfoot[C]{\textbf{\thepage}}
}

\begin{document}

\pagestyle{SPstyle}

\begin{center}{\Large \textbf{\color{scipostdeepblue}{
%%%%%%%%%% TODO: Write your article's title here
Electron charge dynamics and charge separation: A response theory approach \\
}}}\end{center}

\begin{center}\textbf{
% Use (full) first name (+ middle name initials) + surname format.
% Separate subsequent authors by a comma, omit comma and use "and" for the last author.
% Mark the corresponding author(s) with a superscript symbol in this order
% \star, \dagger, \ddagger, \circ, \S, \P, \parallel, ...
Lionel Lacombe\textsuperscript{1,2}
Lucia Reining\textsuperscript{1,2} and
Vitaly Gorelov\textsuperscript{1,2$\star$},
}\end{center}

\begin{center}
{\bf 1}
LSI, CNRS, CEA/DRF/IRAMIS, École Polytechnique, Institut Polytechnique de Paris, F-91120 Palaiseau, France
\\
{\bf 2} European Theoretical Spectroscopy Facility (ETSF)
%%%%%%%%%% END TODO: AFFILIATIONS
% Provide email address of corresponding author(s)
\\[\baselineskip]
$\star$ \href{mailto:vitaly.gorelov@polytechnique.edu}{\small vitaly.gorelov@polytechnique.edu}
\end{center}

\section*{\color{scipostdeepblue}{Abstract}}
\textbf{\boldmath{%
	This study applies response theory to investigate electron charge dynamics, with a particular focus on charge separation. We analytically assess the strengths and limitations of linear and quadratic response theories in describing charge density and current, illustrated by a model that simulates charge transfer systems. While linear response accurately captures optical properties, the quadratic response contains the minimal ingredients required to describe charge dynamics and separation. Notably, it closely matches exact time propagation results in some regime that we identify. We propose and test several approximations to the quadratic response and explore the influence of higher-order terms and the effect of an on-site interaction $U$. }}
	%Additionally, we analyze the role of Coulomb interactions and observe that under strong correlation, the crossover from linear to quadratic regimes occurs at higher perturbation strengths. Our findings demonstrate the potential of response theory up to second order for modeling charge dynamics in complex structures over a range of timescales.

%The abstract is in boldface, and should fit in 8 lines. It should be written in a clear and accessible style, emphasizing the context, the problem(s) studied, the methods used, the results obtained, the conclusions reached, and the outlook. You can add a table contents, recommended if your paper is more than 6 pages long.

\vspace{\baselineskip}

%%%%%%%%%% BLOCK: Copyright information
% This block will be filled during the proof stage, and finilized just before publication.
% It exists here only as a placeholder, and should not be modified by authors.
\noindent\textcolor{white!90!black}{%
\fbox{\parbox{0.975\linewidth}{%
\textcolor{white!40!black}{\begin{tabular}{lr}%
  \begin{minipage}{0.6\textwidth}%
    {\small Copyright attribution to authors. \newline
    This work is a submission to SciPost Physics. \newline
    License information to appear upon publication. \newline
    Publication information to appear upon publication.}
  \end{minipage} & \begin{minipage}{0.4\textwidth}
    {\small Received Date \newline Accepted Date \newline Published Date}%
  \end{minipage}
\end{tabular}}
}}
}
%%%%%%%%%% BLOCK: Copyright information

% For convenience during refereeing we turn on line numbers:
%\linenumbers

\vspace{10pt}
\noindent\rule{\textwidth}{1pt}
\tableofcontents
\noindent\rule{\textwidth}{1pt}
\vspace{10pt}

% Write your article contents here, starting from first \section.
% An example structure is given below.

\section{Introduction}

Accurate modeling of electron charge dynamics is crucial for addressing a wide range of fundamental questions in physics and chemistry. Understanding the mechanisms of charge transfer and the separation of electrons and holes is a key challenge. These processes play a pivotal role in numerous technological applications, including photovoltaics, photocatalysis, and chemical reactions. For example, the efficient generation and transfer of charge carriers directly impacts the design and performance of solar cells, while charge separation dynamics are fundamental to catalyzing chemical transformations in many photo-activated processes. 

To model these phenomena, various computational approaches have been developed, each with its own set of advantages and limitations. Among the most widely used techniques are Time-Dependent Density Functional Theory (TDDFT) \cite{Ullrich2012,marques2012fundamentals} and Non-Equilibrium Green’s Functions (NEGF) \cite{Stefanucci2010}. 

Real-time implementation of TDDFT to describe excited charge dynamics \cite{Fuks2013,Schaffhauser2016,Maitra2017} is favored for its balance of computational efficiency and accuracy. However, it still struggles with incorporating many-body effects, since the exact exchange-correlation (xc) functional is unknown \cite{Lacombe2023}, and it is still computationally challenging to access long time scales. Current approximations to the xc functional cannot accurately describe the charge transfer, although a partial description of charge transfer is possible in generalized Kohn-Sham, for example using hybrid functionals \cite{Maitra2017,Kummel2017}. There has been a recent progress, with a reformulation of TDDFT that uses response quantities for real time propagation \cite{Dar2024}.

\textit{Ab-initio} calculations using NEGF are possible under the Generalised Kadanoff-Baym Anzatz \cite{GKBA1986}. This robust approximation offers a rigorous treatment of electron-electron and electron-phonon interactions at non-equilibrium conditions. Although it can scale linearly with propagation time \cite{Pavlyukh2022}, it remains computationally demanding and it is limited to short time scales, small systems, or parameterized model Hamiltonians \cite{Dahlen2007,Attaccalite2011,Latini2014,Miguel2016,Pavlyukh2021,Pavlyukh20222,Tuovinen2023,Perfetto2023,Sangalli2021,Pavlyukh2024,Stefanucci2025,Caruso2025, Reeves2025}. 

A central challenge in modeling charge dynamics is the development of theoretical frameworks that remain accurate across time scales, perturbation strengths, and interaction regimes. This work explores a general response theory approach, including both linear \cite{Kubo1957} and quadratic \cite{Boyd2020} formulations, to compute time-dependent observables. Linear response is widely used for calculating optical properties such as absorption and electron energy loss \cite{Onida2002}, while quadratic response captures nonlinear phenomena like second-harmonic generation \cite{Shen, Boyd2020}. Quadratic response also enables the study of nonlinear effects such as the electro-optic effect \cite{Prussel2018} and the shift current responsible for the bulk photovoltaic effect \cite{Cheng2021}. However they are rarely used to describe charge dynamics.

In practice, first- and second-order response functions are often computed within the independent-particle approximation, sometimes accounting for local field effects \cite{Luppi2010,Luppi2010b,Hubener2010,Attaccalite2013}, and extended to include excitonic effects using the Bethe–Salpeter equation \cite{Hubener2011,Attaccalite2011,Riefer2017}. 

However, while response theory allows systematic inclusion of many-body effects and in principle enables  access to very long simulation times, it remains perturbative and is valid near equilibrium. The goal of this work is to identify the regimes where linear and second-order responses yield reliable results, both on the basis of general arguments and considering our particular case of charge separation in a small model.% in the general case and for a particular systems, where perturbation is applied on one side and charge is propagated to another.

This article will explore the following key questions on charge dynamics and response behavior: \textit{How does charge dynamics differ in linear and quadratic response?} 
\textit{Can charge dynamics and charge separation be accurately described within these two frameworks? What is the range of validity for linear and quadratic response in relation to perturbation amplitude, time- and length- scales? And finally, how does the Coulomb interaction influence these findings?}

The structure of the article is as follows. Section~\ref{sec:timeprop} introduces the exact time propagation from the ground state. Linear and quadratic response formalisms for time-dependent observables such as charge density and current are presented in Sec.~\ref{sec:response}, with higher-order contributions discussed in Sec.~\ref{sec:chiN}. A minimal site-based model, mimicking an optoelectronic device, is introduced in Sec.~\ref{sec:model}. The results are analyzed in Sec.~\ref{sec:results}, beginning with the validity of linear and quadratic regimes for charge density (Sec.~\ref{sec:chi1}) and current (Sec.~\ref{sec:current}). Section~\ref{sec:chi2analys} provides an analysis of perturbation amplitude and the crossover from linear to higher-order regimes across observables, followed by Section~\ref{sec:chi2} with practical approximations to second-order response. Section~\ref{sec:interaction} demonstrates the effect of an onsite interaction $U$ on the accuracy of the response approach. Conclusions are presented in Sec.~\ref{sec:conclusion}.

\section{Theoretical framework}

\subsection{Time propagation}\label{sec:timeprop}

We consider a system subject to a time-dependent perturbation added to a static Hamiltonian $\hat{H} + \hat{V}_{ext}(t)$, where the scalar potential $\hat{V}_{ext}(t) = \int d\mathbf{r} \hat{n}(\mathbf{r}, t) V_{ext}(\mathbf{r}, t)$ couples to the electron density operator $\hat{n}(\mathbf{r}, t)$. Other types of external perturbations, such as vector electric or magnetic fields, would simply lead to modifications of the perturbation matrix elements ($V_{IJ}$, see Eq.~\ref{eq:VIJ}).
%We choose a scalar perturbation here in order to express the system’s dynamics in terms of the familiar response functions, which can be computed using the Bethe–Salpeter equation within the framework of many-body perturbation theory \cite{Onida2002}. 
A many-body state $|\Psi(t) \rangle$ evolves in time as: 
\bea
|\Psi(t) \rangle = \mathcal{T} \exp \left( -i [ \hat{H} t + \int_{-\infty}^t dt' \hat{V}_{ext}(t')] \right) | \Psi_0 \rangle,
\label{eq:timeprop}
\eea
where $| \Psi_0 \rangle = | \Psi(t=0) \rangle$ and the evolution operator is the time-ordered exponential with $\cal{T}$ as the time-ordering operator  \cite{Stefanucci2010}.
We implement this expression by discretizing time into small time intervals, $dt$.  At each time step $s$, the operator $\hat{H} + \hat{V}_{ext}(s dt)$ is diagonalized, such that $\hat{H} + \hat{V}_{ext}(s dt) = UDU^{-1}$, where $D(s)$ is the diagonal matrix of eigenvalues and $U(s)$ is the matrix of eigenvectors. The exponential operator can then be expressed as  $ e^{-i dt (\hat{H} + \hat{V}_{ext}(s dt))} =U e^{-i dt D} U^{-1}$. Albeit restricted to small systems, this approach has the benefit of being easy to implement and the propagation remains unitary. 
%\lionel{There is also the notion of eigenfrequencies being closer to the exact ones than a Crank-Nicolson approach for example. It is an issue I faced in the past and I read something on the subject (Crank Nicolson eigenfrequencies are actually the eigenvalues of arctan(Hdt) and not Hdt) but I cannot find the article to cite.}
With decreasing $dt$, the result converges to the exact solution.

The time evolution of observables, specifically, the electron density $\hat{O} \equiv \hat{n}(\mathbf{r})$ and current density $\hat{O} \equiv \hat{\mathbf{j}}(\mathbf{r})$, is analyzed in this work. Expanding the time-dependent wave function in the eigenstates of the unperturbed Hamiltonian, $\hat{H}$, $| \Psi(t) \rangle = \sum_I e^{-iE_It} c_I(t) | \Psi_I \rangle $,
a time-dependent observable $O(t)$ can be expressed as

\beq
O(t) = \langle \Psi(t) | \hat{O} | \Psi(t) \rangle = \sum_{IJ} e^{i\Delta_{IJ}t} c^*_I(t) c_J(t) O_{IJ} \equiv \sum_{IJ} \rho_{JI}(t) O_{IJ},
\label{eq:timepropbasis}
\eeq
where $\Delta_{IJ} = E_I - E_J$ denotes the energy difference between eigenstates, and $O_{IJ} = \langle \Psi_I | \hat{O} | \Psi_J \rangle$ is the matrix element of the observable, which governs the spatial structure of the charge dynamics. The many-body  density matrix $\rho_{JI}(t) =  e^{i\Delta_{IJ}t} c_I^*(t) c_J(t)$ encodes the quantum coherence of the time-dependent state $| \Psi(t) \rangle$ \cite{Boyd2020,Baumgratz2014}. 
The time-dependent coefficients $c_I(t)$ evolve according to

\bea
\label{eq:timeprop_basis}
c_I(t) = \sum_{J}  \left[ \exp \left(-i\int_{-\infty}^t \hat{V}_{ext}(t') dt' \right) \right]_{IJ} c_J(0),
\eea
where evolution operator in interaction picture is the time ordered exponential. When the initial state is the ground state, $c_J(0) = c_0(0) \delta_{J0}$ and the sum over $J$ disappears. 
The matrix elements of $\hat{V}_{ext}(t)$ , 
\beq
\label{eq:VIJ}
V_{IJ}(t) =  e^{i\Delta_{IJ}t} \int d\rv V_{ext}(\rv,t) n_{IJ}(\rvec)
\eeq
are non zero only when the state $I$ and state $J$ have a non-vanishing spacial overlap between each other and with the perturbation $V_{ext}(\rv,t)$.

\subsection{Response theory}\label{sec:response}

Response theory is obtained by expanding the time-ordered exponential in Eq. \ref{eq:timeprop_basis} in powers of the perturbation $\hat{V}_{ext}(t)$. This expansion breaks unitarity, leading to a violation of wavefunction normalization; truncating at $n$-th order introduces errors of order $n+1$ in the perturbation strength. Applying this expansion to the time-dependent density matrix yields \\
$
\rho_{IJ}(t) = \sum_{n=0}^{\infty}\rho^{(n)}_{IJ}(t).
$
The general form of the response theory involves time-ordered nested commutators of the perturbation with the unperturbed Hamiltonian,

\bea
\rho^{(n)}_{IJ}(t) = e^{-i\Delta_{IJ}t} (-i)^n \int_{-\infty}^{t} dt_1 ... \int_{-\infty}^{t_{n-1}} dt_n  [\hat{V}_{ext}(t_1),[\hat{V}_{ext}(t_2)...,[\hat{V}_{ext}(t_n),\hat{\rho}^{(0)}]...]]_{IJ}.
\label{eq:fullDM}
\eea
Starting from the ground state, the zeroth order density matrix is $\rho^{(0)}_{IJ} = \delta_{I0}\delta_{J0}$.

Traditionally, the system’s response to external perturbations is expressed in terms of response functions \cite{Kubo1957,Fetter1971}. Up to second order, the change in the expectation value of an observable $\hat{O}$, defined as $\delta \langle \hat{O}(t) \rangle = \langle \hat{O}(t) \rangle - \langle \hat{O}(0) \rangle$, induced from the ground state by a scalar external potential $V_{ext}(\mathbf{r}, t)$, can be written in terms of the corresponding linear and nonlinear response functions.

\bea
\delta \langle \hat{O}(\rv,t) \rangle = && \int d\rv_1 dt_1 \chi_{\hat{O}}^{(1)} (\rv,\rv_1,t-t_1) V_{ext}(\rv_1,t_1) \\ \nonumber  &+& \int d\rv_1 dt_1 d\rv_2 dt_2 \chi_{\hat{O}}^{(2)}(\rv,\rv_1,\rv_2,t-t_1,t-t_2) V_{ext}(\rv_1,t_1) V_{ext}(\rv_2,t_2) + ... .
\label{eq:response}
\eea
Throughout this article we will focus on space dependent observables, such as charge and current densities, since we are interested in charge dynamics and separation.
The first $\chi_{\hat{O}}^{(1)}$ and the second $\chi_{\hat{O}}^{(2)}$ order response functions are the response of an observable $\hat{O}$ to an external potential $V_{ext}(\rv,t)$. %\vitaly{This comes later: In the following, we will discuss in detail contributions from linear and quadratic response for the case of the dynamics of charge density $n(\rv,t)$ and current density $\mathbf{j}(\rv,t)$.}  

One key strength of response theory is its ability to efficiently incorporate interactions, including excitons via the Bethe–Salpeter equation \cite{Onida2002} and electron–phonon or plasmon couplings through, for example, cumulant expansions \cite{Cudazzo2020,Cudazzo2020a}.

%A key strength of response theory is its ability to efficiently incorporate interactions such as electron–hole and electron–phonon coupling \cite{Onida2002,Giustino2017}. It is possible to include many-body effects, such as excitonic interactions, using the Bethe-Salpeter equation \cite{Onida2002}, moreover, one can include into the electronic response function coupling to bosons such as plasmons and phonons in the framework of cumulant expansion \cite{Cudazzo2020,Cudazzo2020a}. 
Moreover, unlike exact time propagation, which requires re-solving the full dynamics for each perturbation, response functions can be computed once to be reused later for different perturbations. This facilitates efficient evaluation of system responses to arbitrary perturbations, time profiles, and field strengths, of course within the limits of validity of perturbation theory.

In practice, response functions are often approximated to reduce computational complexity. Common approaches include the random phase approximation (RPA), time-dependent Hartree-Fock (TDHF) and approximations to time-dependent Green's function functional theory (TDGFFT), such as time-dependent GW (TDGW) \cite{Perfetto2022}, and approximations to time-dependent density functional theory (TDDFT) \cite{Onida2002,Ullrich2012}. These methods introduce varying degrees of approximation to the exchange and correlation effects. In this work, the focus is on the exact expressions for the response functions, derived directly from the time-dependent many-body wavefunction, to provide a benchmark for assessing the accuracy of response theories, and we will mostly look at the non-interacting problem.

\subsubsection{Linear response and its limitations}

The analysis begins with the dynamics obtained from linear response, which follows from Eq.~\ref{eq:fullDM} with $n=1$. %The analysis begins with the density response, followed by the current. 
Throughout, only systems with real-valued wavefunctions $\Psi_I$ are considered.
In that case, the linear response function for any time- and space- dependent observable $\hat{O}(\rv,t)$, $\chi_{\hat{O}}^{(1)}$, can be expressed as

\bea
\chi_{\hat{O}}^{(1)}(\rv,\rv_1; t-t_1) = -2\theta(t-t_1) \sum_{I} O_{0I}(\rv)n_{I0}(\rv_1) A [\Delta_{I0}(t-t_1)] %\sin[\Delta_{I0}(t-t_1)],
\label{eq:chi1}
\eea
where $\theta(t)$ is a Heaviside step function, $O_{IJ}(\rv) = \langle \Psi_I | \hat{O}(\rv) | \Psi_J \rangle$, $A[\cdot] \equiv \sin[\cdot]$ for a real observable, for which $\hat{O}^*=\hat{O}$, and $A[\cdot] \equiv i \cos[\cdot]$ for an imaginary observable operator, for which $\hat{O}^*=-\hat{O}$. %Note that the diagonal elements $n_{00}$ do not contribute to the summation.
Assuming that the external perturbation can be factorized into spatial and temporal components $V_{ext}(\rv,t) = u(\rv) e(t)$, the first order response reads 
\beq
\delta O^{(1)}(\rv,t) = -2 \sum_I O_{0I}(\rv) \tilde{V}_{I0} \int_{-\infty}^{t} dt_1 A_{I0}(t-t_1) e(t_1),
\label{eq:chi1obs}
\eeq
%\vitaly{what about an observable that doesn't depend on r?}
where $\tilde{V}_{IJ} = \int d\rv_1 u(\rv_1) n_{IJ}(\rv_1)$ is the spatial overlap of the states with the perturbation.

\begin{figure*}[th]
\center
\begin{minipage}[b]{\columnwidth}
\center
       \begin{minipage}[b]{0.9\columnwidth}
        \includegraphics[width=\columnwidth]{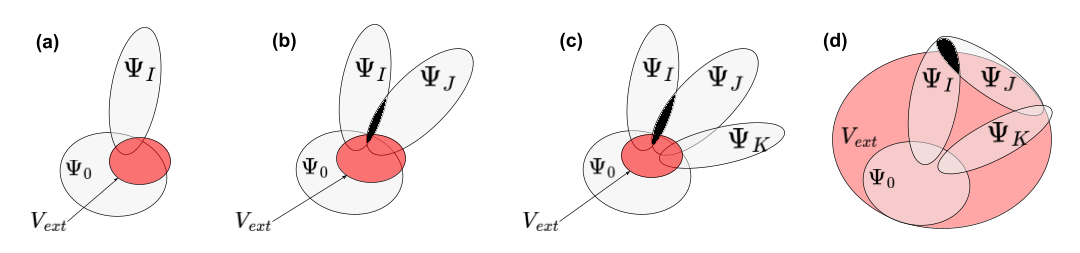}
        \end{minipage}
\end{minipage}
	\caption{\small{Schematic illustration of spacial overlaps between the excited states $\Psi_I$, the external perturbation $V_{ext}$ and the ground state $\Psi_0$ that contribute in response theory up to the first order (a) (see Eqs.~\ref{eq:chi1obs},\ref{eq:dm1}), second order (b) (see Eqs.~\ref{eq:rho2sin_final},\ref{eq:dm2}) and third order (c), (d) (see Eq.~\ref{eq:dm3}) for only one element $O_{IJ}$ of the sum in Eq.~\ref{eq:timepropbasis}. (c) for a localized perturbation in third order (d) for a delocalized perturbation in third order. Black shade is the region, where the observable matrix element $O_{IJ}$ can be non-zero .}
	\label{fig:overlap}}
\end{figure*}

Comparison of Eq.~\ref{eq:chi1obs} with the exact time propagation in Eq.~\ref{eq:timepropbasis} reveals that first-order response lacks terms involving excited-state couplings, $O_{IJ}(\mathbf{r})$ for $I,J \neq 0$. Since it depends only on the overlap between the ground state, an excited state and the perturbation, the linear response is not able to have a non-vanishing value beyond the extention of the ground state (see Fig.~\ref{fig:overlap}(a) for illustration), unless the observable operator is non-local.  % and $\mathbf{j}_{IJ}(\mathbf{r})$.

This means that, when the ground state is mostly localised in the region where the perturbation is applied, a charge transfer is only possible between the states overlapping with the ground state (see Fig.~\ref{fig:overlap}(a) for illustration). The response at distances far from the localisation of the ground state cannot be observed. The current operator involves the gradient, which is not purely local operator, and results in a slightly more delocalized expectation value than the charge density. Therefore, it favours a slightly better linear response result in the region where the ground state has very low amplitude. This will be further illustrated in Section~\ref{sec:results}.

The time-dependent contribution to the response changes depending on the observable.
For a real observable, such as the induced charge density $\delta n(\rv,t)$, and a local, instantaneous perturbation of the form $V_{ext}(\mathbf{r}, t) = u(\rv) \delta(t)$, the time-dependent part of Eq.~\ref{eq:chi1obs} becomes
\beq
\theta(t)\int_{-\infty}^{t} dt_1 A_{I0}(t-t_1) \delta(t_1) = \theta(t)\sin[\Delta_{I0} t].
\label{eq:chi1delta}
\eeq
The response is governed by oscillations at frequencies corresponding to energy differences $\Delta_{I0}$. 

In the case of an oscillating perturbation in time, $V_{ext}(\rv,t) = u(\rv) \theta(t) \sin(\omega t)$, the corresponding contribution to the linear induced charge density becomes

\beq
\theta(t)\int_{-\infty}^{t} dt_1 A_{I0}(t-t_1) \theta(t_1) \sin(\omega t_1) = \theta(t)\frac{\Delta_{I0} \sin[\omega t]- \omega \sin[\Delta_{I0}t] }{\Delta_{I0}^2-\omega^2}.
\label{eq:chi1sin}
\eeq
At resonance, where $\Delta_{I0} = \omega$, the charge propagation oscillates as \\ $(\sin[\omega t] - \omega t \cos[\omega t])/2\omega$. 

Charge separation can be intuitively understood as the condition where the induced density increases on one side of the system and decreases on the other. For positive induced density one can define the electron propagation and for negative, the hole propagation. 

For both perturbations, charge separation can only occur on time scales shorter than $t < \pi/\Delta_{I0}$ (if $\omega>\Delta_{I0}$ for oscillating perturbation, otherwise shorter than $t < \pi/\omega$), past this time, the charge density is changing sign. 
For both considered perturbations, the induced charge would always oscillate around zero in every point in space, indicating that there is no possibility to see a net charge separation in linear response %\vitaly{with these perturbations?} 
(see section~\ref{sec:chi1} for an illustration). 

Considering now an imaginary operator, such as the current-density for which $\hat{\mathbf{j}}^{*}(\rv) = -\hat{\mathbf{j}}(\rv)$. 
%where $\mathbf{j}_{0I}(\rv) = \langle \Psi_0 | \hat{\mathbf{j}}(\rv) | \Psi_i \rangle$. 
In the case of an instantaneous  perturbation, the current density can be obtained by replacing $\sin[\Delta_{I0}t]$ by $i\cos[\Delta_{I0}t]$ in Eq.~\ref{eq:chi1delta}.  For a real valued wavefunction $\mathbf{j}_{00}(\rv) = 0$, meaning that the time average of linear response current is zero, no DC component is present.

In the case of an oscillating perturbation, $e(t) = \theta(t) \sin(\omega t)$, the resulting time-dependent part of the current density is given by

\bea
\label{eq:curr0sin_final}
\theta(t) \int_{-\infty}^{t} dt_1 A_{I0}(t-t_1) \theta(t_1) \sin(\omega t_1) = \theta(t) i \frac{2\Delta_{I0}(\cos[\Delta_{I0} t] - \cos[\omega t]) }{\omega^2-\Delta_{I0}^2}.
\eea
At the resonance $\Delta_{I0} = \omega$ the time dependency is governed by $t \sin(\omega t)$. A crucial condition for observing net charge transport is the presence of a nonzero DC component in the current, i.e., a finite time-averaged value $\langle \mathbf{j}(\mathbf{r}, t) \rangle_t \neq 0$. Such a contribution is absent in the non-resonant linear response regime, where the current oscillates symmetrically around zero.
However, averaging the time-dependent part over a period of oscillations, $T = 2\pi/\omega$, at the resonance results in $2\pi/\omega^2$ and, depending on the corresponding matrix elements in Eq.~\ref{eq:chi1obs}, might be non-vanishing resulting in a finite DC component. 

For both considered perturbations and observables, the ground state doesn't contribute to the summation in Eq.~\ref{eq:chi1obs}, as $\Delta_{00}=0$ and $\mathbf{j}_{00}(\rv) = 0$. Any linear combination of perturbations leads to a corresponding linear superposition of the individual linear responses.

\subsubsection{Quadratic response: analysis and limitations}\label{sec:quadratic} 

Charge separation and propagation beyond the region of localization of the ground state requires the quadratic response $\chi^{(2)}$, defined as:

\bea
 \chi_{\hat{O}}^{(2)}(\rv,\rv_1,\rv_2,t,t_1,t_2) =&& -2\theta(t-t_1)\theta(t-t_2) \theta(t_1-t_2) \Big[ \sum_{IJ} \\
&&  O_{0I}(\rv)n_{IJ}(\rv_1)n_{J0}(\rv_2)  A\left[\Delta_{I0}t+\Delta_{JI}t_1+\Delta_{0J}t_2\right] \nonumber \\ 
&& -O_{IJ}(\rv)n_{0I}(\rv_1)n_{J0}(\rv_2)  A\left[\Delta_{I0}t_1+\Delta_{JI}t+\Delta_{0J}t_2\right]  \Big],  \nonumber
\label{eq:chi2real}
\eea
%\vitaly{where the time ordering operator $T$ is acting on two time variables and is defined as \\ $Tf(t_1,t_2) = \frac12 [\theta(t_1-t_2) f(t_1,t_2) + \theta(t_2-t_1) f(t_2,t_1)]$} and  
where $A[\cdot] \equiv \sin[\cdot]$ for a real observable and $A[\cdot] \equiv i \cos[\cdot]$ for an imaginary observable operator.
%\eea
A second order contribution to an observable $\hat{O}$ in response to a generic perturbation is
\bea
\label{eq:rho2sin_final}
\delta O^{(2)}(\rv,t) &=& 2 \sum_{J \neq 0} (O_{00}(\rv)-O_{JJ}(\rv)) \tilde{V}_{0J}^2 B^{'}_{0J}(t) \nonumber \\ % \frac{ 2\omega^2 +  \omega(\Delta_{J0}-\omega)\cos[(\Delta_{J0}+\omega)t] - \omega (\Delta_{J0}+\omega)\cos[(\Delta_{J0}-\omega)t]}{(\Delta_{J0}^2-\omega^2)^2} \nonumber  \\ 
&& - 2 \sum_{I>J, J \neq 0} O_{IJ}(\rv) \tilde{V}_{0I}\tilde{V}_{J0}  B^{''}_{IJ}(t) \nonumber \\
&& + 2 \sum_{I \neq 0,J \neq 0} O_{0I}(\rv) \tilde{V}_{IJ}\tilde{V}_{J0} B^{'}_{IJ}(t),
\eea
where the terms have been regrouped by their importance for charge dynamics. %This is a main result of this article. 
The first term is the dominant one. It comes from expanding the time-dependent wavefunctions (or coefficients) up to the first order of Eq.~\ref{eq:timepropbasis}, only keeping the diagonal contributions, and it is determined by the excited state expectation values of the observable. These quantities can be obtained from linear response TDDFT \cite{Furche2002,Flores2014}. % multiplied by a constant (for density) or linear in time (for current) term comming from the $B^{'}_{0J}(t)$ as will be shown later. 
The second term, arising from transitions between excited states, originates from the second-order expansion of the time-dependent wavefunctions in Eq.~\ref{eq:timepropbasis}. The third term, in the case when $O_{0I}(\mathbf{r}) \approx 0$ outside the ground state localization, contributes little to the second-order response, this is the linear-response-like term. %The spatial part of the summand in the second term is symmetric under the exchange of indices $I \leftrightarrow J$ this is why the summation is restricted to only $I>J$ terms. 
Unlike the linear response, the first two terms in the second-order contribution involve the full set of matrix elements $O_{IJ}(\mathbf{r})$, qualitatively they contain all the information of the exact time-propagation expression in Eq.~\ref{eq:timepropbasis}. Relying on the excited states overlaps, the second order is able to propagate charge beyond the extension of the ground state, see the black shade on Fig.~\ref{fig:overlap}(b). However, the overlap between an excited state, the ground state and the perturbation, $\tilde{V}_{0I}$, has to be non-zero.

Evaluating Eq.~\ref{eq:rho2sin_final} requires computing the overlap integrals $\tilde{V}_{IJ}$, which can be performed in advance. Importantly, only the elements involving the ground state of the $\tilde{V}_{IJ}$ matrix enter the first two terms of Eqs.~\ref{eq:rho2sin_final}. %The third term in both expressions can be neglected for a particular system considered here.
Considering the symmetry of the spacial part of the second term in Eq.~\ref{eq:rho2sin_final} with respect to the exchange of $I$ and $J$, the sum is reduced to only terms with $I>J$. The temporal dependence is governed by the factors $B^{'}$ and $B^{''}$. 

In the case of the second order induced charge density $\delta n(\rv,t)$ for the instantaneous perturbation in time, $V_{ext}(\rv,t) = u(\rv) \delta(t)$, the time-dependent parts $B^{'}_{IJ}(t)$ and $B^{''}_{IJ}(t)$ become:

\bea
\label{eq:rho2delta}
&B^{'}_{IJ}(t) = \theta(t) \int_0^{t} dt_1 \int_0^{t_1} dt_2 \cos\left[\Delta_{I0}t+\Delta_{JI}t_1+\Delta_{0J}t_2\right]  \delta(t_1)  \delta(t_2) = \theta(t) \cos\left[\Delta_{I0}(t)\right] \\ \nonumber
&B^{''}_{IJ}(t) =\theta(t) \int_0^{t} dt_1 \int_0^{t_1} dt_2 \cos\left[\Delta_{I0}t_1+\Delta_{JI}t+\Delta_{0J}t_2\right]  \delta(t_1)  \delta(t_2) = \theta(t) 2\cos\left[\Delta_{JI}(t)\right],
\eea
which results in $B^{'}_{0J}(t)=\theta(t)$. This gives a constant offset to the charge dynamics necessary to the charge separation.

Considering further an oscillating time dependent part of the external potential, \\ $V_{ext}(\rv,t) = u(\rv) \theta(t) \sin(\omega t)$, the second order time integrals become:

\bea
\label{eq:chi2time}
B^{'}_{IJ}(t) &=& \theta(t)\int_0^{t} dt_1 \int_0^{t_1} dt_2 \cos\left[\Delta_{I0}t+\Delta_{JI}t_1+\Delta_{0J}t_2\right]  \sin(\omega t_1)  \sin(\omega t_2) =  \nonumber \\
&=& \frac{\theta(t)}{2(\Delta_{J0}^2-\omega^2)} \Big(\frac{2\omega^2 \cos[\Delta_{I0}t] + \omega (\Delta_{JI}-\omega)\cos[(\Delta_{J0}+\omega)t] - \omega (\Delta_{JI}+\omega)\cos[(\Delta_{J0}-\omega)t]}{(\Delta_{JI}^2-\omega^2)} - \nonumber \\
&& - \frac{\Delta_{0J}}{\Delta_{I0}} (\cos[\Delta_{I0}t] - 1) + \frac{(\Delta_{J0}\Delta_{I0} + 2\omega^2)(\cos[2\omega t]-\cos[\Delta_{I0}t])}{(\Delta_{I0}^2-4\omega^2)} \Big), 
\eea
\bea
B^{'}_{0J}(t) = \theta(t) \Big[ \frac{(1-\cos[2\omega t])}{4(\Delta_{J0}^2-\omega^2)}+\frac{2\omega^2 + \omega (\Delta_{J0}-\omega)\cos[(\Delta_{J0}+\omega)t] - \omega (\Delta_{J0}+\omega)\cos[(\Delta_{J0}-\omega)t]}{2(\Delta_{J0}^2-\omega^2)^2} \Big],
\eea
\bea
\label{eq:intIJ}
B^{''}_{IJ}(t) &=& \frac{\theta(t)}{2(\Delta_{J0}^2-\omega^2)(\Delta_{I0}^2-\omega^2)} \Big( 2\omega^2 \cos[\Delta_{IJ}t] + (\Delta_{J0}\Delta_{I0} + \omega^2) - (\Delta_{J0}\Delta_{I0} - \omega^2) \cos[2 \omega t] \nonumber \\
&&- \omega (\Delta_{I0}+\omega)\cos[(\Delta_{J0}-\omega)t] - \omega (\Delta_{J0}+\omega)\cos[(\Delta_{I0}-\omega)t] \nonumber \\
&& + \omega (\Delta_{I0}-\omega)\cos[(\Delta_{J0}+\omega)t] + \omega (\Delta_{J0}-\omega)\cos[(\Delta_{I0}+\omega)t] \Big).
\eea
In $B^{'}_{IJ}(t)$ the resonance is reached when either $\omega$ or $2\omega$ match the transition energies.
The $B^{''}_{IJ}(t)$ term is resonant only when the transition energies match $\omega$. 
The time dependence of the second-order response features oscillations at frequencies determined by the transition energies $\Delta_{IJ}$, the perturbation frequency $\omega$ and $2\omega$, as well as their sums and differences. Beyond the oscillatory behavior, a key contribution arises from the constant offset in $B_{0J}^{'}(t)$, which ensures that the induced density vanishes at $t = 0$. This term shifts the dynamics to enable charge separation (see Sec.~\ref{sec:results}). 
At resonance conditions $\Delta_{J0/I0} \pm \omega = 0$, the second order response diverges faster than the first order (due to the $1/(\Delta_{J0}^2-\omega^2)^2$ in $B_{0J}^{'}(t)$), indicating a breakdown of the response theory. Additional resonances occur when $\Delta_{I0} \pm 2\omega = 0$, but these are less significant far from the ground-state region, since it only enters the third term of Eq.~\ref{eq:rho2sin_final}.

In the second order current-density response, for real valued wavefunctions $\mathbf{j}_{II}(\rv) = 0$, the first term in Eq.~\ref{eq:chi2real} disappears. For an instantaneous delta-function perturbation, the current density follows from Eq.~\ref{eq:rho2delta} by replacing $\cos[\Delta_{JI/I0}t]$ with $i \sin[\Delta_{JI/I0}t]$. In the case of an oscillating perturbation, the corresponding expression for the current density time integrals that enter Eq.~\ref{eq:rho2sin_final} read:

\bea
\label{eq:curr_time}
B^{'}_{IJ}(t) 
&=& \frac{\theta(t)}{2(\Delta_{J0}^2-\omega^2)} \Big(\frac{2\omega^2 \sin[\Delta_{I0}t] + \omega (\Delta_{JI}-\omega)\sin[(\Delta_{J0}+\omega)t] - \omega (\Delta_{JI}+\omega)\sin[(\Delta_{J0}-\omega)t]}{(\Delta_{JI}^2-\omega^2)} - \nonumber \\
&& - \frac{\Delta_{0J}}{\Delta_{I0}} \sin[\Delta_{I0}t] - \frac{(\Delta_{J0}\Delta_{I0} + 2\omega^2)\sin[\Delta_{I0}t] - 2\omega(\Delta_{I0} + 2\Delta_{J0})\sin[2\omega t] }{(\Delta_{I0}^2-4\omega^2)} \Big),
\eea

\bea
\label{eq:curr_time2}
B^{''}_{IJ}(t) 
&=& \frac{\theta(t)}{2(\Delta_{J0}^2-\omega^2)(\Delta_{I0}^2-\omega^2)} \Big(2\omega \Delta_{JI} \sin[2 \omega t] + 2 \omega^2 \sin[\Delta_{JI} t] + \\ \nonumber
&& + \omega (\Delta_{I0}-\omega)\sin[(\Delta_{J0}+\omega)t] - \omega (\Delta_{I0}+\omega)\sin[(\Delta_{J0}-\omega)t] - \\ \nonumber
&&- \omega (\Delta_{J0}-\omega)\sin[(\Delta_{I0}+\omega)t] + \omega (\Delta_{J0}+\omega)\sin[(\Delta_{I0}-\omega)t] \Big). 
\eea
For an oscillating perturbation a non-vanishing DC component in the current emerges at resonance due to the terms like $\lim_{\Delta_{J0} \to \pm \omega} \frac{\sin[(\Delta_{J0} \pm \omega)t]}{(\Delta_{J0} \pm \omega)} = t$. 
In the case of a perturbation by a vector potential, this effect corresponds to the generation of a shift current, also known as the bulk photovoltaic effect (BPVE), as discussed in Refs.~\cite{Sipe2000, Young2012}. This mechanism enables net current flow in the absence of external bias and plays a central role in nonlinear optical and transport phenomena.

In the linear and quadratic response, for an oscillating perturbation, a convergence parameter can be identified from Eqs.~\ref{eq:chi1sin} and \ref{eq:curr0sin_final} in linear and Eqs.~\ref{eq:chi2time}-\ref{eq:curr_time2} in quadratic response, $I/(\Delta_{J0} \pm \omega)$, where $I$ is the perturbation amplitude. The response theory breaks down if $I$ is large or if the resonance is reached $\Delta_{J0} = \pm \omega$.

\subsection{Higher orders: what can they add?}\label{sec:chiN}

To understand the contributions of higher orders, it is useful to look at the density matrix expansion defined in Eq.~\ref{eq:fullDM} and the corresponding overlaps between the states. The focus here will be on propagating the charge far from the ground state and from the perturbation.

The main contributions to the density matrix in first order are coming from one of the elements of the commutator:

\bea
\rho^{(1)}_{IJ}(t) \propto \tilde{V}_{0I} B_{IJ}(t) \delta_{0J}.
\label{eq:dm1}
\eea
In linear response, only off-diagonal elements of the density matrix contribute, while diagonal terms such as $\rho^{(0)}_{00}$ vanish. %As a result, the dynamics are restricted to coherent charge oscillations. This distinction between coherence and incoherence, linked respectively to first- and second-order terms in the time-dependent density matrix, is discussed in Ref.~\cite{Sangalli2018}. 
As a result, net charge transport requires the inclusion of higher-order processes. 
In the second-order response, all matrix elements can contribute, since relevant terms in the commutator of Eq.~\ref{eq:fullDM} are not restricted by $\delta_{0J}$, enabling asymmetry and transport in the induced dynamics. One of the elements of the second order of Eq.~\ref{eq:fullDM} is
\bea
\rho^{(2)}_{IJ}(t) \propto \tilde{V}_{0I} \tilde{V}_{J0} B_{IJ}(t).
\label{eq:dm2}
\eea
The condition for a nonzero second-order contribution is a finite overlap between the excited states $I$ and $J$ that enters the observable $O_{IJ}$ in Eq.~\ref{eq:timepropbasis}, the ground state, and the applied perturbation. Similarly, the third-order (and higher-order) response generate the full set of density matrix elements, e.g.
\bea
\rho^{(3)}_{IJ}(t) \propto \sum_{K} \tilde{V}_{0I} \tilde{V}_{JK}\tilde{V}_{K0} B_{IJK}(t).
\label{eq:dm3}
\eea
In this expression, contrary to the second order, not all the excited states (e.g. $J$) must overlap with the ground state. Instead there is an extra overlap between states $K$ and $J$ and the perturbation and $K$ must overlap with the ground state. %However, it modifies the density matrix only quantitatively, no new matrix elements $\tilde{V}_{IJ}$ are introduced that would qualitatively change the structure at higher orders.

Consider two scenarios: first, when the perturbation is localized in the region of the ground state, which is illustrated in Fig.~\ref{fig:overlap}(c). In this case, the third-order response behaves similarly to the second-order response in terms of charge propagation. Shaded in black area is the overlap between the states $I$ and $J$ that contributes to an observable of interests, $O_{IJ}(\rv)$ (see Eq.~\ref{eq:timepropbasis}). This area is the same in second, third and higher orders, meaning that the third (and higher) order response cannot propagate charge significantly further than the second order for a localized perturbation due to the persistent presence of the $\tilde{V}_{0I}$ terms across all orders.

In the second scenario, involving a fully delocalized perturbation (Fig.~\ref{fig:overlap}(d)), charge transport can extend beyond the range achievable in the second order. Here, when state $J$ overlaps with $K$, which in turn overlaps with the ground state, there is a potential for charge to propagate further (black shaded area) that is not accessible by second order (see Appendix~\ref{app:vfull}).

For the fourth and higher orders, the situation remains similar. For a perturbation localized within the ground state, the requirement remains that all states must overlap with the ground state. The only quantitative change will be added from more elements like $\tilde{V}_{KL}$.  In the case of a completely delocalized perturbation, higher-order responses allow for further charge propagation through sequential overlaps between the states, i.e. there will be additional elements added into the chain, e.g. $\Psi_0$ - $\Psi_I$ - $\Psi_J$ - $\Psi_K$ - $\Psi_0$. However, of course, for a weak perturbation, these contributions will be smaller than in the second order. %Thus, if we perturb the system in one region and observe the response in another, higher-order responses will modify the overall structure but will not introduce new mechanisms to transport charge further away.

\section{The model}\label{sec:model}

\begin{figure*}[th]
\center
\begin{minipage}[b]{\columnwidth}
\center
       \begin{minipage}[b]{0.48\columnwidth}
        \includegraphics[width=\columnwidth]{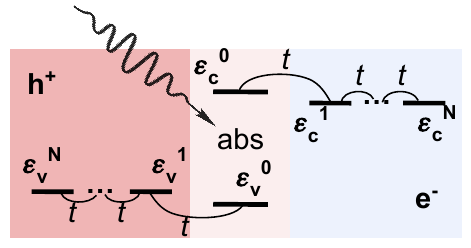}
        \end{minipage}
       \begin{minipage}[b]{0.48\columnwidth}
        \includegraphics[width=\columnwidth]{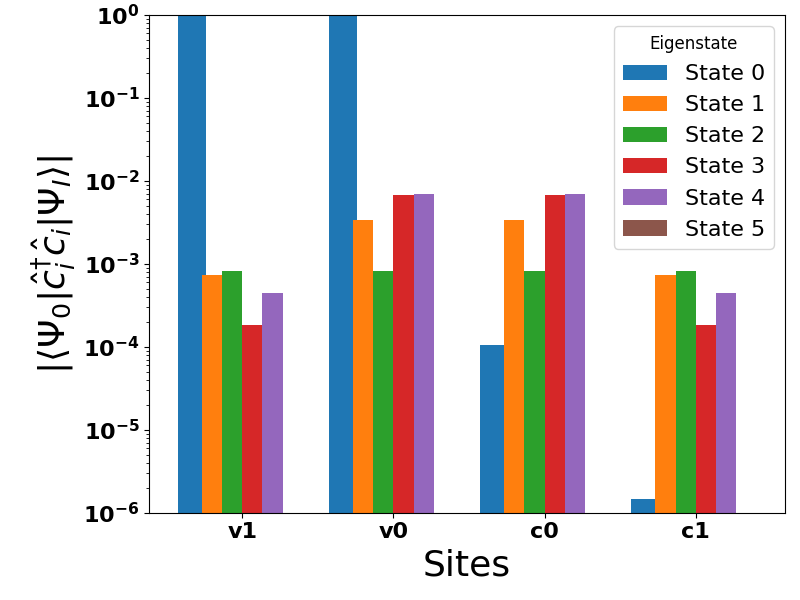}
        \end{minipage}
\end{minipage}
	\caption{\small{Left: Site model used in this work. The model is divided into three regions: absorber in the middle, electron transport layer (blue) and hole transport layer (red). $N+1$ electrons occupy lower energy (valence) sites $vi$ connected with the hopping $t_h=1$. The $N+1$ higher energy (conduction) sites $ci$, connected by a hopping $t_h=1$ are separated by a single site energy $\epsilon^0_c-\epsilon^0_v=10t_h$ in the absorber. The external field arrives in the absorber region. Right: matrix element $\langle \Psi_0 | \hat{c}^{\dagger}_i\hat{c}_i | \Psi_I \rangle$ between the ground and excited states for $t_a=0.1t_h$ projected onto the sites for a system with two valence and two conduction sites. State 5 doesn't have an overlap with the ground State 0.}   
\label{fig:model}}
\end{figure*}

%\vitaly{cite some justification of the model}

To illustrate this discussion and to get further insight, we will use a model that simulates systems with interfaces where an external perturbation excites a charge in one region or material, after which the excited charge propagates into another region or material. This model aims to demonstrate charge separation into electrons and holes, a process prevalent in optoelectronics and particularly in photovoltaic systems. %Systems with interfaces are not centrosymmetric and have both first and second order components of the response.

Our Hubbard-like model consists of $N+1$ lower energy sites and $N+1$ higher energy sites and a number of electrons $N_e=N+1$. The illustration of the model and the extension of the ground state can be found in Fig.~\ref{fig:model}. 
\beq
\hat{H}=\sum_{i=0}^{N} \epsilon_v^i \hat{c}_{vi}^{\dagger} \hat{c}_{vi}+ \sum_{i=0}^{N} \epsilon_c^i \hat{c}_{ci}^{\dagger} \hat{c}_{ci} - \sum_{i=0}^{N-1} t_h (\hat{c}_{vi}^{\dagger} \hat{c}_{vi+1} +\hat{c}_{vi+1}^{\dagger} \hat{c}_{vi}) - \sum_{i=0}^{N-1} t_h (\hat{c}_{ci}^{\dagger} \hat{c}_{ci+1} +\hat{c}_{ci+1}^{\dagger} \hat{c}_{ci}) + t_a (\hat{c}_{v0}^{\dagger} \hat{c}_{c0} +\hat{c}_{c0}^{\dagger} \hat{c}_{v0})
\label{eq:model}
\eeq

The first lower energy site $v0$ and the first higher energy site $c0$ will represent the region, where the external perturbation acts. This region is called absorber. The rest of the low energy sites represent the hole transport layer and the rest of the high energy sites - the electron transport layer. In the following we choose the hopping between the $v0$ and $c0$ sites, $t_a < t_h$, such that, the $N_e=N+1$ electrons in the many-body ground state will mostly occupy $N+1$ lower sites, $vi$, and have a little occupation of higher energy sites, $ci$ (see the blue column in the right panel of Fig.~\ref{fig:model} for a system with 4 sites). Therefore will can call the lower energy sites - valence sites and higher energy sites - conduction sites. No spin is included, except in the last section \ref{sec:interaction}. 

For most of the discussion below we consider 2 valence and 2 conduction sites (unless it is stated otherwise). The characteristic energy difference in our system between conduction and valence sites is $\epsilon_c^0-\epsilon_v^0 = 10t_h$. The energy difference between the absorber region and electron/hole transport layer is $\epsilon_{v/c}^1-\epsilon_{v/c}^0=1t_h$. Except for the validity limits of response theory, the overall conclusions do not depend on the particular parameters.

The external perturbation acts only in the absorber region and is defined as \\ $\hat{V}_{ext}^{ij}(t)= I\delta_{ic0}\delta_{jv0} (\hat{c}^{\dagger}_{c0} \hat{c}_{v0}+\hat{c}^{\dagger}_{v0} \hat{c}_{c0}) e(t)$. We have chosen such perturbation in order to study the dynamics far from it. We have verified that a delocalized perturbation that acts on the whole model doesn't change the main conclusions about the validity of the second order response theory. In both first- and second-order responses, the dominant terms involve $\tilde{V}_{0I}$, it contains the overlaps between excited states and the ground state (see Eq.\ref{eq:chi1obs} and first two terms of Eq.\ref{eq:rho2sin_final}). Figure~\ref{fig:model} (right panel) shows $\langle \Psi_0 | \hat{c}^\dagger_i \hat{c}_i | \Psi_I \rangle$ for $t_a = 0.1t_h$, revealing nearly an order-of-magnitude larger overlap at absorption sites ($v0$, $c0$) than in the transport layers ($v1$, $c1$). Extending the perturbation to other sites thus has negligible impact on $\tilde{V}_{0I}$ (see Appendix~\ref{app:vfull}).  %In the following we will consider the external perturbation as $e(t) = \theta(t)\sin[\omega t]$. 
The strength of the perturbation $I$ and the perturbation frequency $\omega$ are the parameters that differentiate between different response regimes. 

Consider a natural electric field scale in the Hubbard model as a field that does work comparable to a hopping $t_h \sim 1$ eV over a distance $a$: $E_0\sim \frac{t_h}{ea} = 10^{10}$ V/m, where $a=1$ $\AA$ and $e$ - an electric charge. A solar intensity on a clear day can be assumed to be $I_s \approx 1000$ W/m$^2$ \cite{astmG173}, converting it into an electric field intensity makes $E_s \sim 10^6$ V/m, which in the Hubbard model translates to $I \sim 10^{-4}t_h$. This is a very weak field, compared to a typical pump-probe experiment, where a pump probe $I \sim (0.1 - 1)t_h$ \cite{PhysRevB.107.064309,PhysRevB.96.235142}. In fact, as can be seen in the next section, at $I \sim 10^{-4}t_h$ one can already observe some second order effect on charge dynamics.

\section{Results and discussion}\label{sec:results}

Here, the focus is on the excited charge dynamics within the presented model via examining the induced density and current. For the site model, this simplifies to the dynamics of a change on a site occupation number, $\delta n_{ii}(t)$, and the current between two sites $\delta j_{i i+1}(t)$  with respect to the ground state. 
%expressed as $\delta \rho_{ii}(t) = \langle \Psi(t) | \hat{c}^{\dagger}_i \hat{c}_i | \Psi(t) \rangle$ and $\delta j_{i i+1}(t) = \langle \Psi(t) | it_h \hat{c}^{\dagger}_i \hat{c}_{i+1} - it_h \hat{c}^{\dagger}_{i+1} \hat{c}_{i} | \Psi(t) \rangle$. 
Our objective is to achieve charge separation, where electrons are transported to the electron transport layer, i.e. positive $\delta n_{ii}(t)$, and holes to the hole transport layer, i.e. negative $\delta n_{ii}(t)$. This should also result in a net DC current, i.e. $\langle j_{i i+1}(t) \rangle_t \neq 0$. %We expect an increase in density or occupation at the electron layer and a decrease in density (indicating the absence of an electron) at the hole transport layer. For efficient separation of electrons and holes, the induced density should be mostly increasing at the electron layer and mostly decreasing at the hole layer. 

In the following the difference between the exact time propagation defined in Section \ref{sec:timeprop} and the linear and quadratic response theory will be examined. 

\subsection{Charge dynamics: linear and quadratic regime} \label{sec:chi1}

\begin{figure*}[th]
\center
\begin{minipage}[b]{\columnwidth}
\center
        \begin{minipage}[b]{0.32\columnwidth}
        \includegraphics[width=\columnwidth]{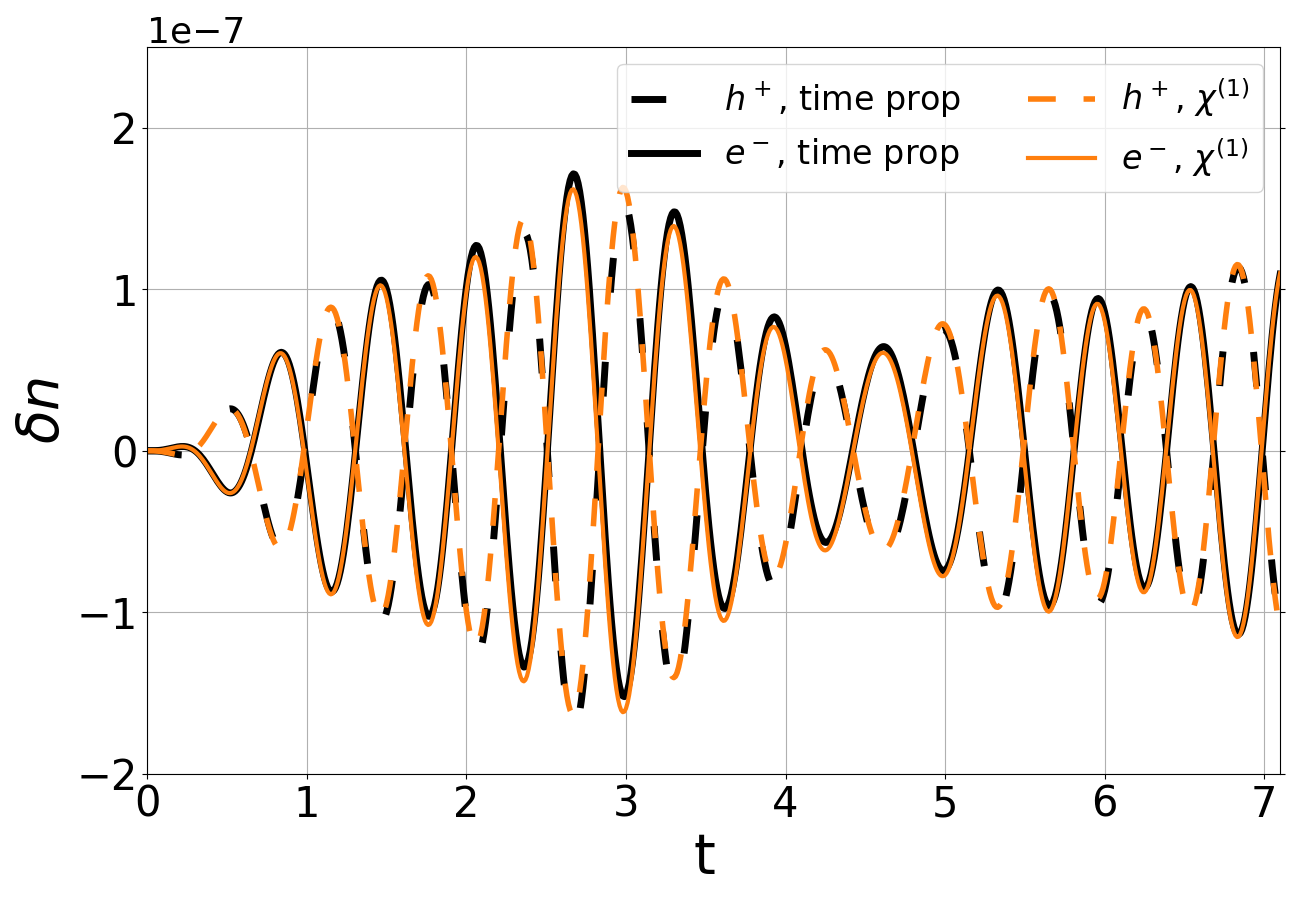}
        \end{minipage}
       \begin{minipage}[b]{0.32\columnwidth}
        \includegraphics[width=\columnwidth]{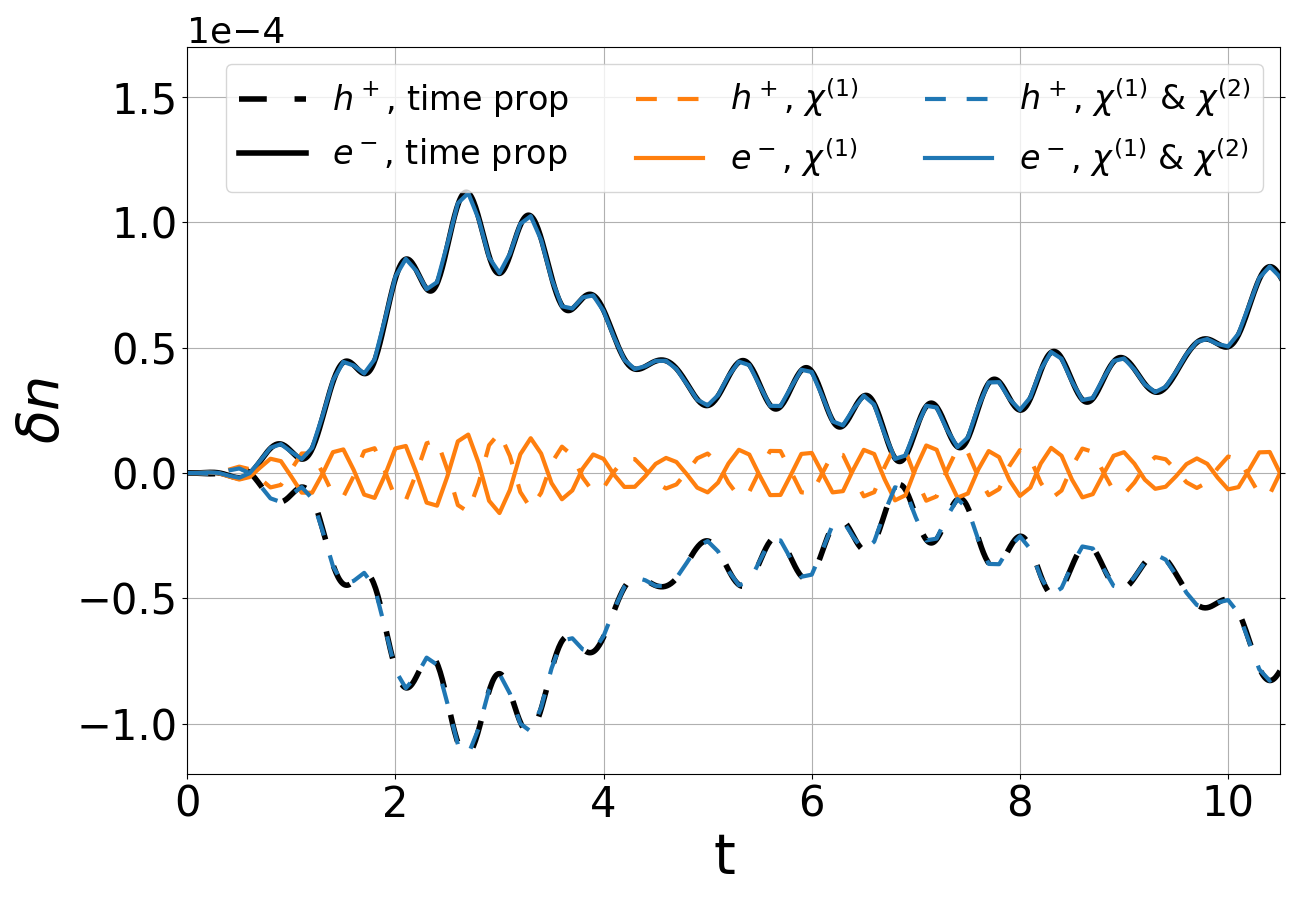}
        \end{minipage}
       \begin{minipage}[b]{0.32\columnwidth}
        \includegraphics[width=\columnwidth]{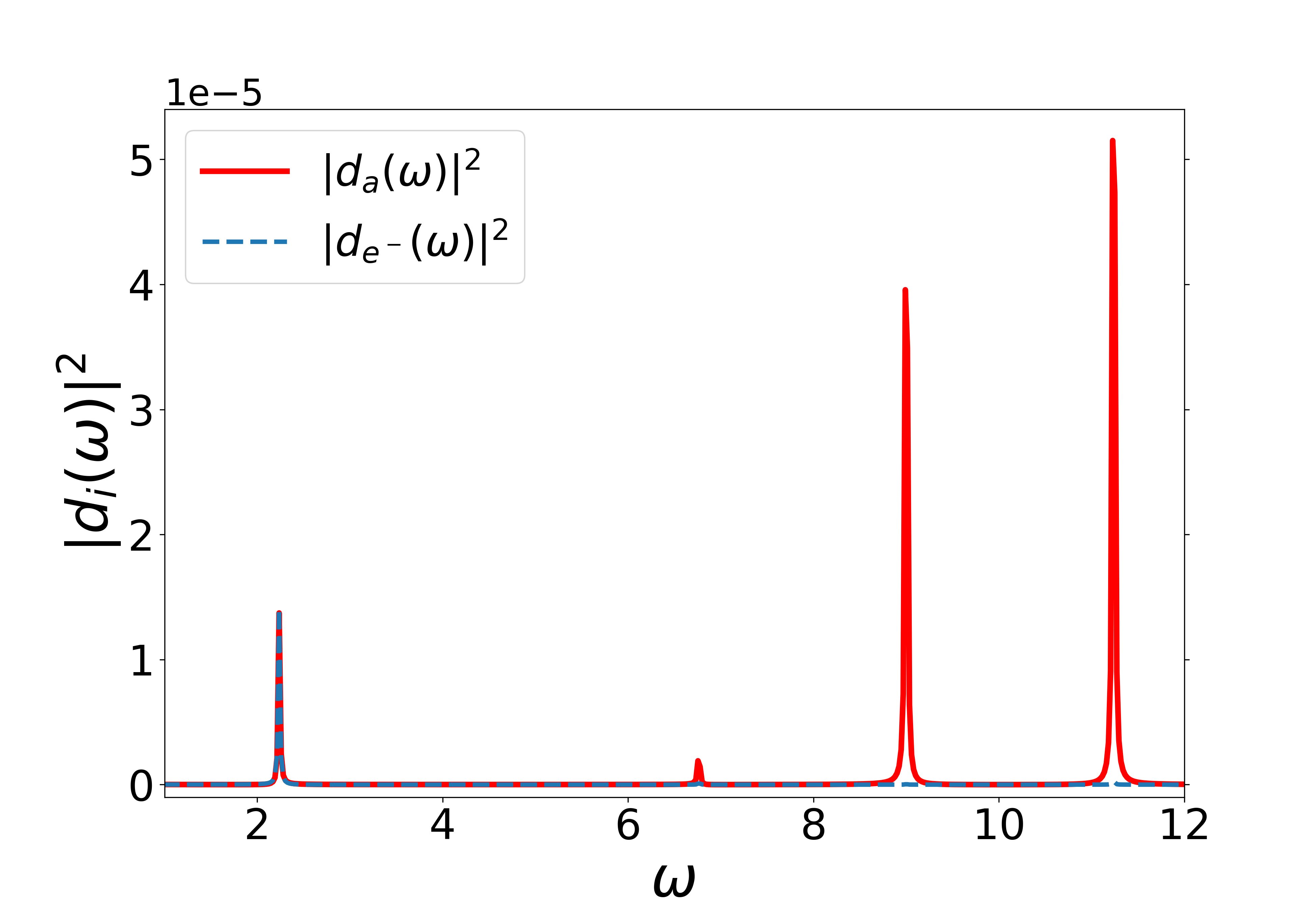}
        \end{minipage}
\end{minipage}
	\caption{\small{ Left and Middle: Density induced by an oscillating perturbation of electron (solid lines) and hole (dashed lines) transport layers for a system of two valence and two conduction sites. Orange lines indicate the linear response contribution. Blue lines - combined linear and quadratic contributions and black lines - exact dynamics. Left: $e(t) = 10^{-4}t_h \cdot \theta(t) \sin(\omega t)$  Middle: $e(t) = 10^{-2}t_h \cdot \theta(t) \sin(\omega t)$. Right: delta-kick perturbation $e(t) = \delta(t)$. Fourier transform of the dipole moment on the absorbtion site $c0$ (red) and on the electron transport layer $c1$ (blue). } 
\label{fig:chi1sin}}
\end{figure*}

Introducing the hopping $t_a \neq 0$ between the valence and conduction sites,  e.g. between site $v0$ and $c0$, the ground state will have contributions from the conduction sites $ci$. Consequently, the matrix element entering the linear response,  $n^i_{0I} \equiv \langle \Psi_0 | \hat{c}_{i}^{\dagger} \hat{c}_{i} | \Psi_I \rangle \neq 0$, will be non-zero (right panel of Fig.~\ref{fig:model}). This results in a non-vanishing linear response contribution, as illustrated by the orange curves in Fig.~\ref{fig:chi1sin} (left and middle panels). Here and throughout the manuscript, the solid lines indicate the charge at electron transport layer and the dashed line the charge at hole transport layer. %When the charge at electron and hole transport layers differs only by a sign, we show only one of the two. 

In this subsection the perturbation frequency is $\omega=10t_h$, which is not resonant. For small enough perturbation $I \leq 10^{-4}t_h$ (Fig. \ref{fig:chi1sin} left), the charge dynamics is predominantly governed by the linear response regime. The linear response (orange lines) closely matches the exact dynamics (black line) over extended simulation periods. Only small differences between the linear response and exact dynamics are observed. 

For perturbations strong enough to separate electrons and holes, $I>10^{-4}t_h$ (see Fig. \ref{fig:chi1sin} middle panel), the linear response induced charge (orange lines) oscillates around zero, as indicated by the time dependency part of Eq.~\ref{eq:chi1sin}. The separation of charge occurs, when the induced density is positive on the electron transport layer (solid lines) and negative on the hole transport layer (dashed lines). The linear response alone always oscillates between positive and negative values and fails to separate charges for any perturbation strength. Including the second order response (blue lines) function $\chi^{(2)}$ recovers completely the charge dynamics. 

For a non-resonant perturbation frequency, no deviation from exact dynamics is observed for any simulation time, which can be explained by the oscillating nature of the response. In fact, according to Eq.~\ref{eq:chi2time}, the time dependency is governed by periodic trigonometric functions. Thus, in the finite system, if the response dynamics is accurate for the largest period of oscillations $t > \frac{2\pi}{\Delta_{JI} \pm \omega}$ (or for the smallest $\Delta_{JI} \pm \omega$ excluding resonance, it remains accurate for any simulation time). %\vitaly{For extended systems, ...}  
As system size increases, the longest oscillation periods grow, making it increasingly challenging for second-order response to reproduce exact dynamics at long times.
%\vitaly{check it} 

Remarkably, we have additionally verified that even up to 20 sites, no deviation is seen between the linear (left of Fig.~\ref{fig:chi1sin}) or quadratic and linear (middle of Fig. \ref{fig:chi1sin}) response and exact time propagation at the most distant sites from the absorber for the considered perturbations. This is again  explained by the fact that, for the particular localised spacial perturbation the higher orders do not add new possibilities to transport charge (see analyses in Section \ref{sec:chiN}).

The right panel of Fig.~\ref{fig:chi1sin} shows the Fourier transform of the dipole within the absorber (orange) and in the electron transport layer (blue) for a delta-kick perturbation, applied locally in the absorber. This perturbation allows us to see the full spectrum of excitations. It illustrates how second-order effects dominate away from the absorber region. While linear response accurately captures coherent dynamics at the perturbation site, relevant for optical absorption, second-order contributions become dominant farther out at electron/hole transport layers, enabling charge transport (see middle panel of Fig.~\ref{fig:chi1sin}). This aligns with Fig.~\ref{fig:overlap}, which shows that second-order matrix elements allow for charge propagation away from the perturbation.

\subsection{Current vs. charge dynamics }\label{sec:current}

\begin{figure*}[th]
\center
\begin{minipage}[b]{\columnwidth}
\center
        \begin{minipage}[b]{0.8\columnwidth}
        \includegraphics[width=\columnwidth]{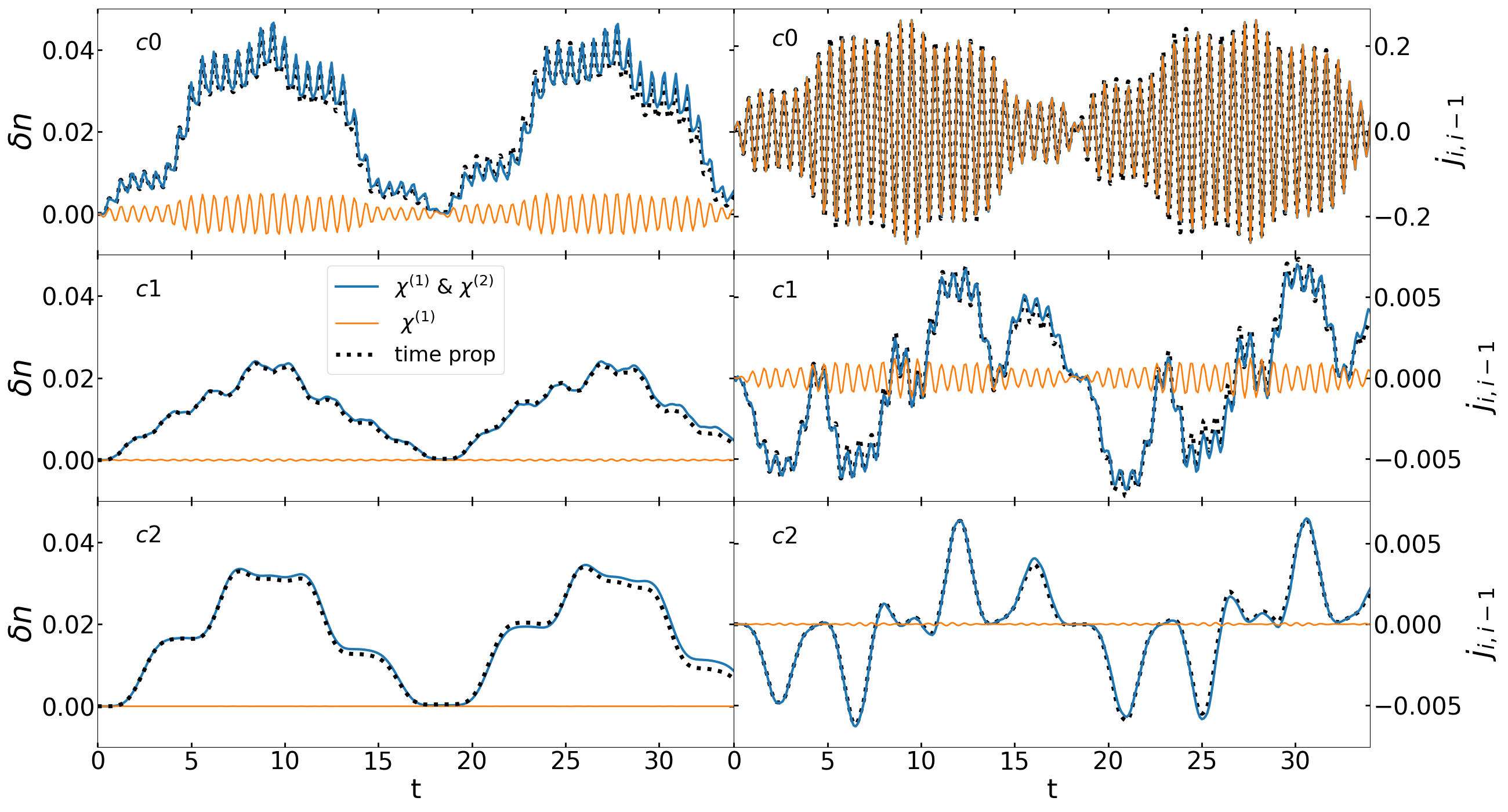}
        \end{minipage}
\end{minipage}
	\caption{\small{Model with 3 valence and 3 conduction sites with an oscillating external perturbation $0.01t_h \sin(\omega t)$ for non-resonant $\omega$. Exact time propagation (black) and the response theory (blue) and only from the linear response (orange). Left: Occupation of 3 conduction sites (top to bottom: c0, c1 and c2). Right: The current between the two sites (v0-c0, c0-c1 and c1-c2). } 
\label{fig:chicurr}}
\end{figure*}

\begin{figure*}[th]
\center
\begin{minipage}[b]{\columnwidth}
\center
        \begin{minipage}[b]{0.9\columnwidth}
        \includegraphics[width=\columnwidth]{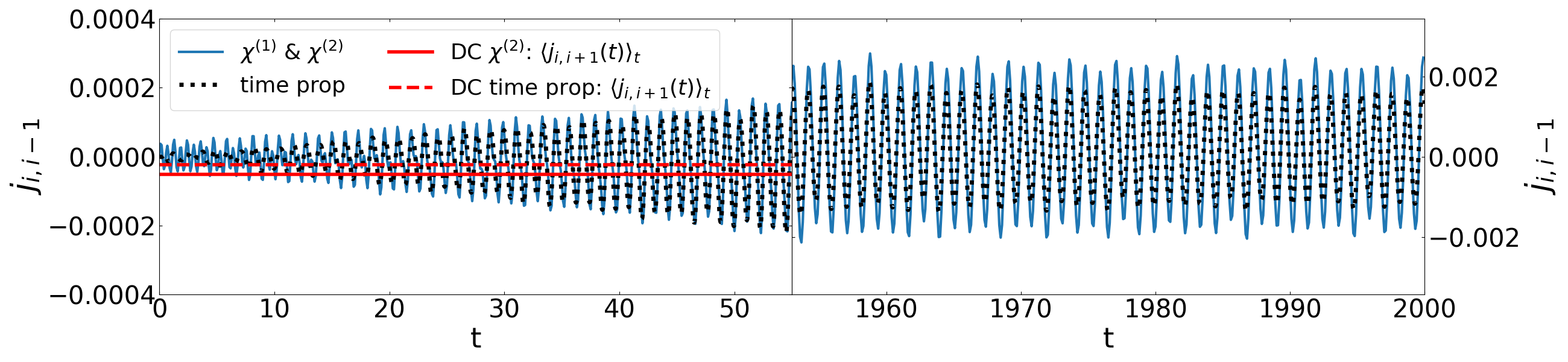}
        \end{minipage}
\end{minipage}
	\caption{\small{Same model as Fig.~\ref{fig:chicurr}. Current at resonance frequency at the last site of electron level $c1-c2$ evaluated in response theory (blue) and in exact time propagation (black). Left: beginning of the dynamics, where the time propagation and second order response agree. Right: later part of the dynamics, where the response starts to diverge. Red lines are the average current over the full time interval (DC component) from second order response (solid) and from time dynamics (dashed). Note that the scale is different between the two panels.} 
\label{fig:chicurr_res}}
\end{figure*}

We will now compare the current density evaluated in linear Eq.~\ref{eq:curr0sin_final} and quadratic Eqs.~\ref{eq:curr_time}-\ref{eq:curr_time2} response to an exact time-dependent current density evaluated with Eq.~\ref{eq:timepropbasis}. The same current can be obtained from the continuity equation, $\frac{\partial \rho(\rv,t)}{\partial t} = - \nabla \cdot \mathbf{j}(\rv,t)$, which is valid order-by-order. 

Fig. \ref{fig:chicurr} shows the density (left panels) and current (right panels) for three conduction sites. The perturbation is oscillating with a non-resonant frequency, the amplitude is such that a slight deviation between the second order response and time propagation can be seen. The advantage of the response theory is that it is possible to separate oscillating linear response from a less oscillating second order. This can be useful during the analysis and for a potential comparison to experimentally measured current. 
In the absorber region $c0$, where the linear response contribution is similar in magnitude to the second-order term, the current, primarily capturing charge oscillations, can be well approximated by the linear response. This is not true for the density itself, so clearly, the validity range of the linear response current extends further than that of the linear response charge density. %In fact, in the charge density we observe an accumulation of charge or an overal shift. 
The current, which is proportional to the time derivative of the charge density, is more sensitive to the oscillations of the density rather then to its overalls shifts. Moreover, the current operator is more delocalized than the density operator, which also results in an extended validity of the linear response current. 
Fig.~\ref{fig:chicurr} also demonstrates the conclusions of section \ref{sec:chiN}: for a localised perturbation, in the linear response one can only propagate charge in the vicinity of the perturbation and close to the ground state. In the second order, where the overlaps between the excited states are present, charge can be propagated much further.  

Figure~\ref{fig:chicurr_res} shows the current at resonance in response theory and in exact propagation. %Interestingly, at the resonance, the response theory current agrees slightly better with the exact time propagation than the density.
The left panel shows the beginning of the simulation (up to $60 t_h$), as in the case of charge dynamics, the two approaches, exact and response, agree (except some oscillations in the beginning that are due to the resonance). 
Even at the end of the simulation the difference between the response theory current and the exact one is not dramatic (right panel of Fig. \ref{fig:chicurr_res}). However, the charge dynamics obtained from response theory deviates from the exact propagation and diverges rapidly (see blue lines in the upper left panel of Fig.~\ref{fig:chi2sintest}), leading to an unphysical total occupation. While the timescale of divergence differs in smaller systems, we have verified that the conclusion holds for the system considered here. The good behavior of the current can be seen by computing the average current (red lines of Fig. \ref{fig:chicurr_res}), also being a DC component of current density or shift current (for this model) can only be obtained at resonance, as can be seen from Eqs.~\ref{eq:curr_time} and~\ref{eq:curr_time2}. Up to $\sim 1000 t_h$ two DC components (from second order and from time propagation) agree, after that the second order DC component starts to diverge from the exact one, and it will continue to diverge due to the presence of a linear in time term in the expression Eq.~\ref{eq:curr_time}.

\subsection{Second order: limits and analyses} \label{sec:chi2analys}

\begin{figure*}[th]
\center
\begin{minipage}[b]{\columnwidth}
\center
       \begin{minipage}[b]{0.44\columnwidth}
        \includegraphics[width=\columnwidth]{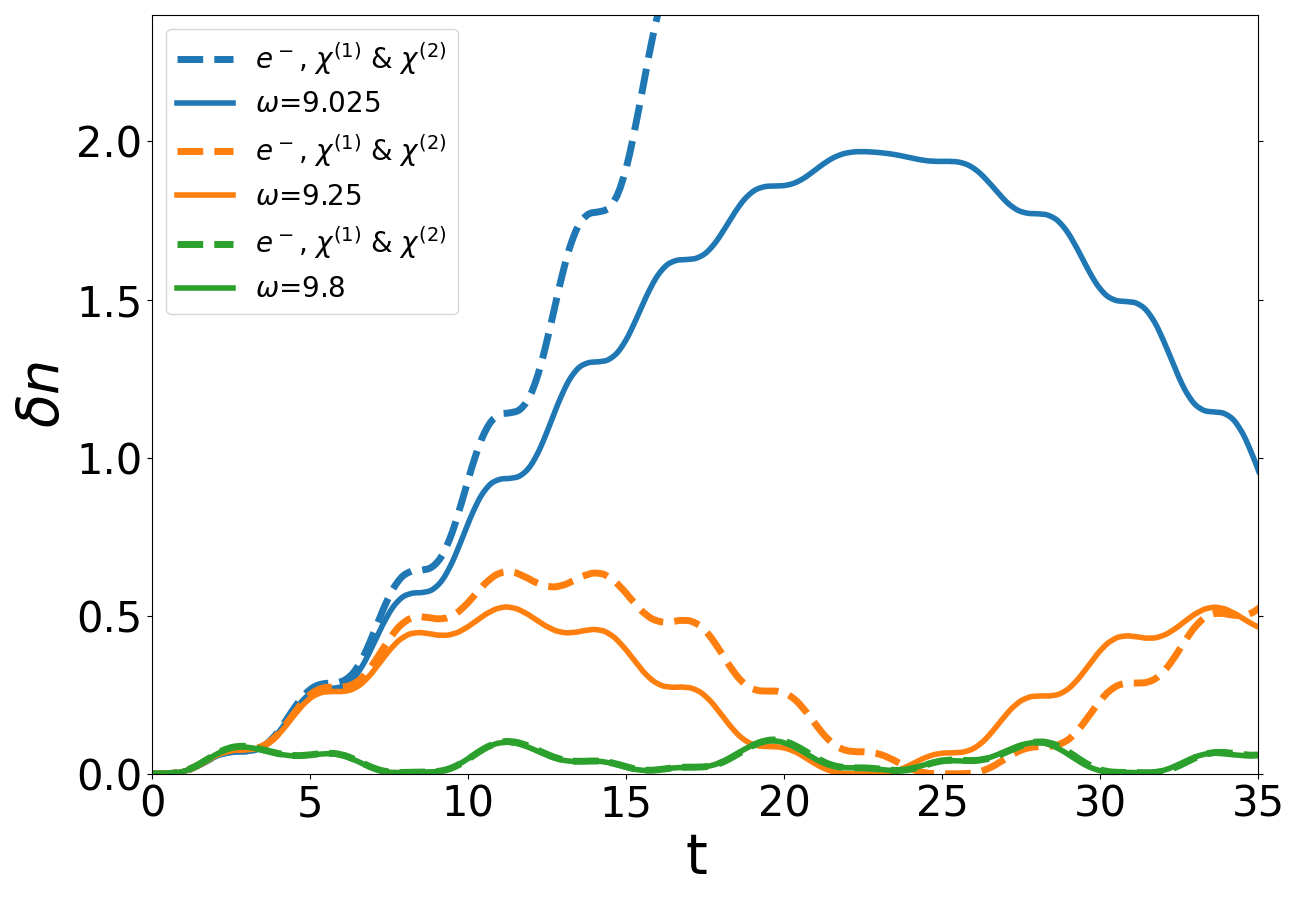}
        \end{minipage}
       \begin{minipage}[b]{0.44\columnwidth}
        \includegraphics[width=\columnwidth]{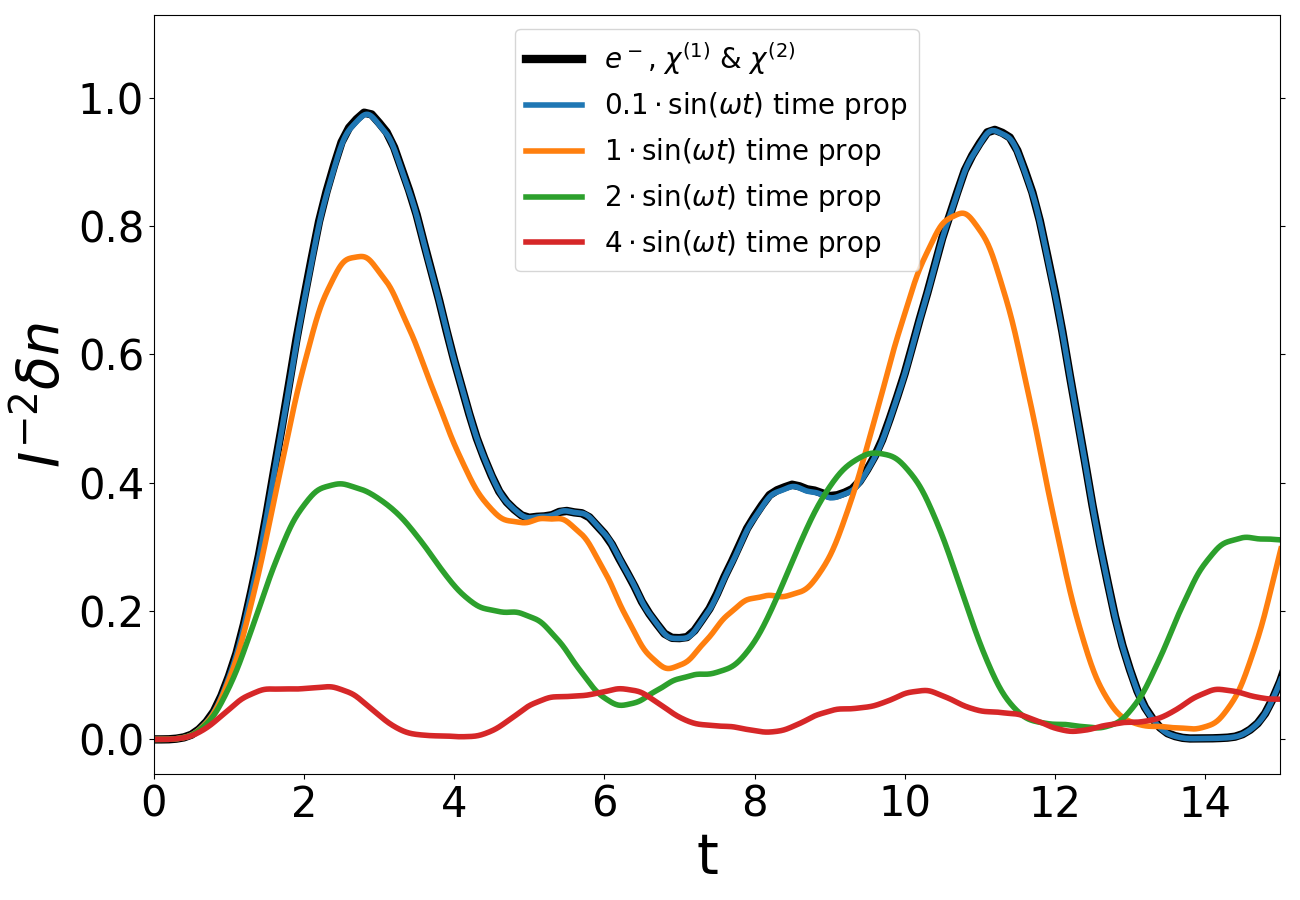}
        \end{minipage}
       \begin{minipage}[b]{0.44\columnwidth}
	\includegraphics[width=\columnwidth]{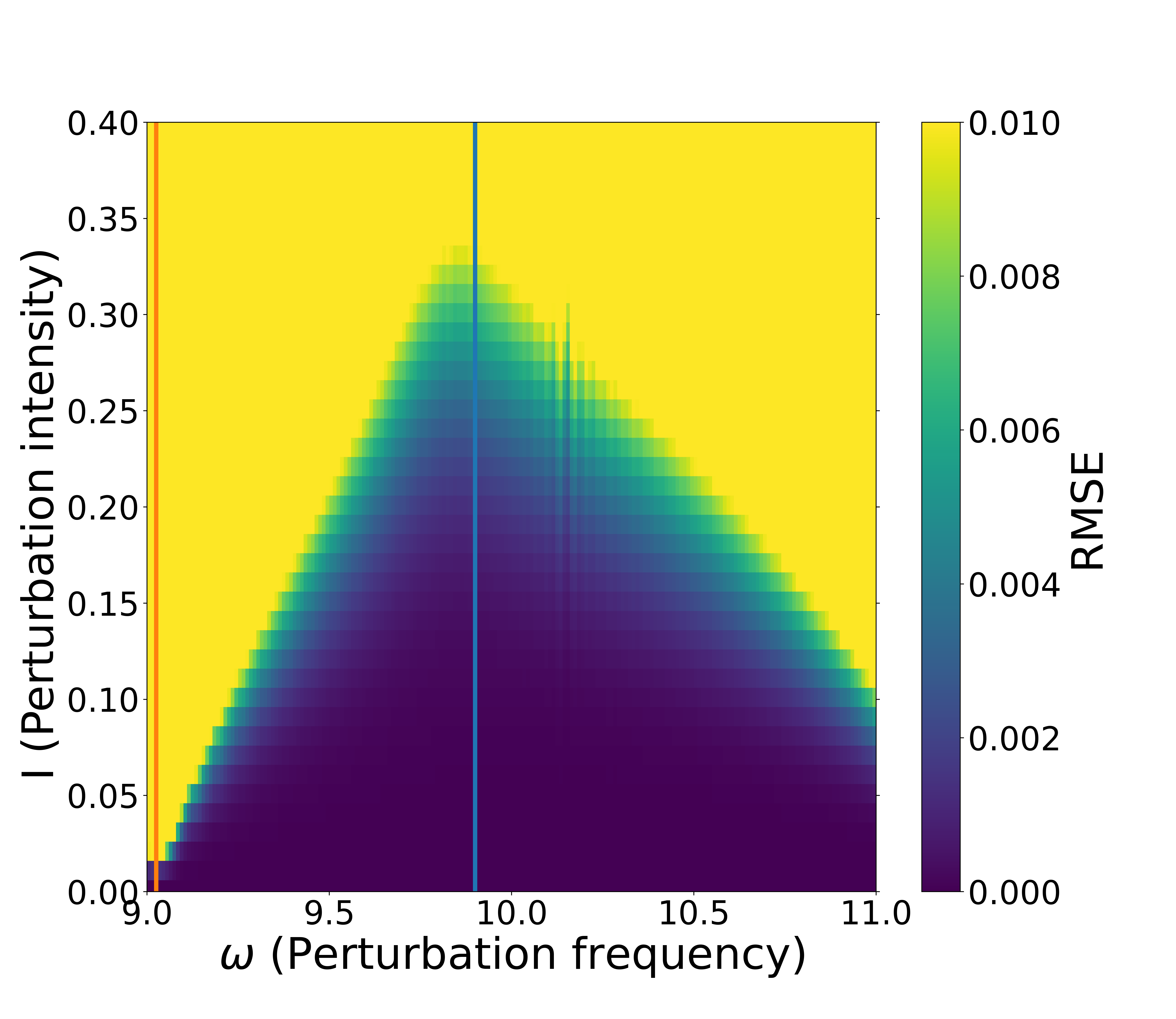}
        \end{minipage}
       \begin{minipage}[b]{0.44\columnwidth}
	\includegraphics[width=\columnwidth]{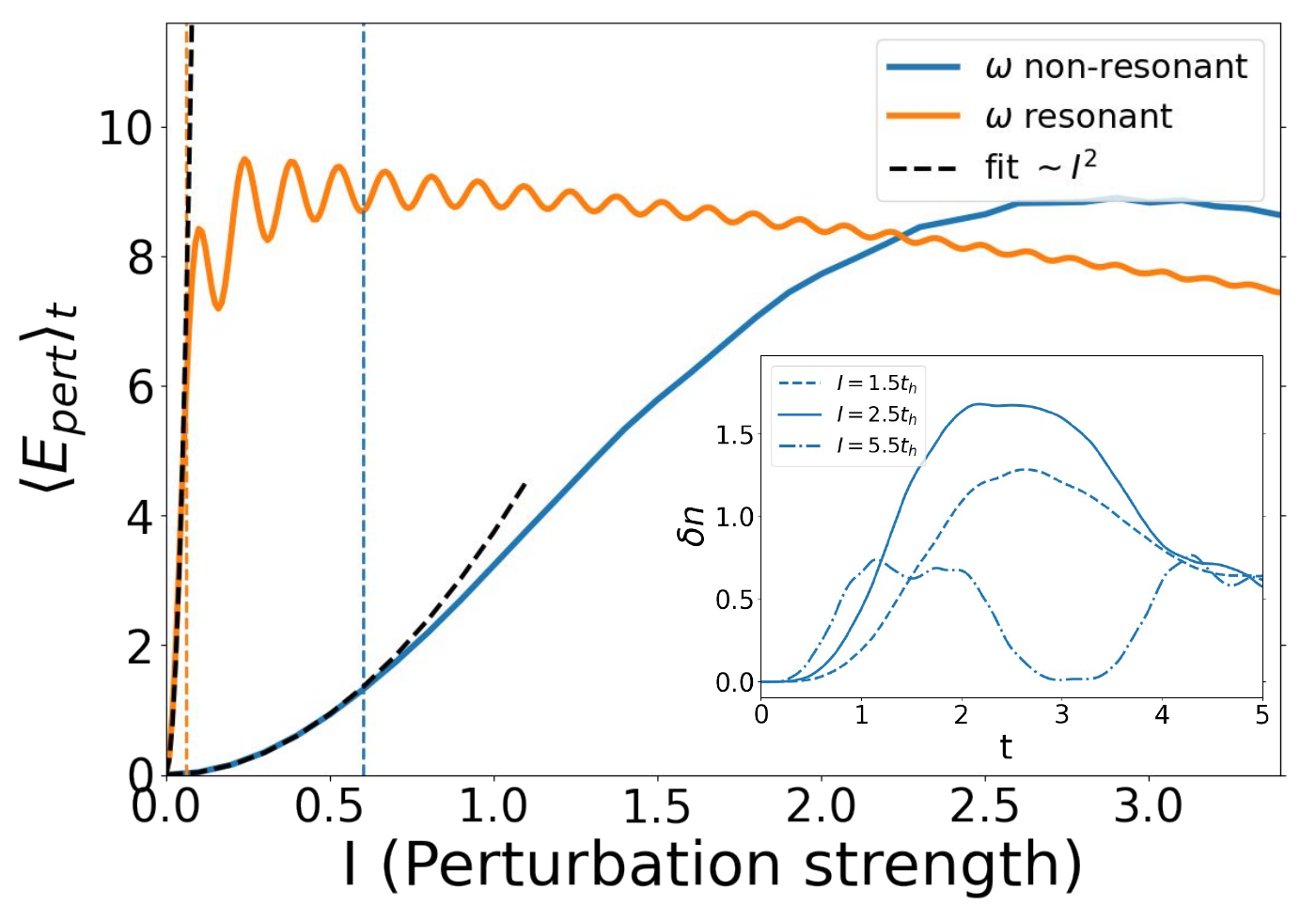}
        \end{minipage}
\end{minipage}
\caption{\small{On the top panels we are following occupation of the electron transport layer. The hopping between the conduction and valence sites is $t_a=0$, which makes the linear response to vanish. The perturbation is $e(t)=I \sin(\omega t)$. 
Top left: Quadratic response and exact time propagation for different perturbation frequencies $\omega$. 
Top right: Black solid line result obtained using $\chi^{(2)}$, colored lines correspond to exact time propagation  $I^{-2} \cdot n^{RT}(t)$. 
Bottom left: Root mean squared error map of the induced density in the electron transport layer obtained up to quadratic response as a function of perturbation frequency $\omega$ and amplitude $I$ (in the units of $t_h$).
Bottom right: Average energy received by the system over a fixed simulation time as a function of perturbation strength obtained from exact time propagation for resonant (orange) and non-resonant (blue) frequencies. Black dashed lines are quadratic fits. Blue and orange dashed vertical lines correspond to the validity range of the quadratic fit. The inset (obtained with exact time propagation) shows the exact charge density at electron layer for a non-resonant perturbation at three different intensities.}
\label{fig:chi2sintest}}
\end{figure*}

Here the limits of quadratic response theory in terms of perturbation frequency and amplitude will be examined. Here and in the next sections only charge density dynamics will be discussed. % we have verified that the conclusions are the same for current density as well. 
First, in Fig.~\ref{fig:chi2sintest} (top left panel), keeping the perturbation amplitude fixed at $I=0.1t_h$ we vary the frequency from the near resonant (blue lines) to a non-resonant (orange line). Out-of-resonance the response aligns with the exact dynamics (green line). Approaching the resonance, the denominator in the second order equations Eqs.~\ref{eq:chi2time}-\ref{eq:intIJ} approaches zero, making the response theory non-convergent (blue and orange dashed lines). 
In the exact time propagation (solid lines) in the resonance ($\Delta_{IJ} = \omega$) the response can be fitted to $\cos[(\Delta_{IJ} \pm \omega_{eff})t]/(\Delta_{IJ} \pm \omega_{eff})$ with an effective frequency $\omega_{eff} < \omega$. The role of higher orders is to introduce such more fluctuating terms.  

Turning to the variations of perturbation amplitude, a fixed off-resonant frequency is considered. As shown in Fig.~\ref{fig:chi2sintest} (top right panel), the quadratic response remains accurate up to $I = 0.3t_h$. Beyond this threshold, higher-order effects become visible: the amplitude of charge oscillations decreases, and additional high-frequency components appear. The bottom left panel of Fig.~\ref{fig:chi2sintest} shows the root mean squared error (RMSE) as a function of external perturbation amplitude and frequency. Using a threshold RMSE of 0.01, the second-order response is valid for $I \lesssim 0.01t_h$ near resonance (orange line) and up to $I \sim 0.3t_h$ in the off-resonant regime (blue line). From the overall trends in the upper panels and in the bottom left panel of Fig.~\ref{fig:chi2sintest}, a convergence parameter for the validity of the response theory becomes evident: $I/\Delta_{IJ} \pm \omega < 1$ that was already discussed at the end of section \ref{sec:quadratic}. Approaching the resonance, the valid perturbation amplitude is decreasing proportionally. This condition delineates the regime where the perturbative expansion converges and second-order theory is quantitatively reliable.

Next, we investigate the amount of energy transferred to the system for a fixed period of time as a function of the perturbation amplitude, defined as:

\beq
\langle E_{pert} \rangle_T = \frac1T \int_T dt \langle \Psi(t) | \hat{H} | \Psi(t) \rangle - E_0. 
\eeq
The bottom right panel of Fig.~\ref{fig:chi2sintest} illustrates this quantity, computed in exact time propagation, for a resonant (orange) and non-resonant (blue) frequency. Characteristic energy behaviour for a linear regime is when the energy delivered to the system by an external field is proportional to the square of its amplitude, $E_{\text{pert}} \propto I^2$,  \cite{Shen}. The figure shows that this holds up to  $I \approx 0.6t_h$ for a non-resonant and up to $I \approx 0.1t_h$ for a resonant frequency. Above this values higher-order effects emerge. These thresholds exceed the intensities where deviations in second-order charge density from exact dynamics become visually apparent (see Fig.~\ref{fig:chi2sintest}, bottom left; orange for resonant, blue for non-resonant cases). This indicates that the energy transferred into the system is still within the linear regime, when the deviations in the second order response appear. The inset of  Fig.~\ref{fig:chi2sintest} (bottom right) shows the charge density at electron layer for a non-resonant perturbation for three intensities, before the energy saturation (dashed), at the maximum of energy (solid) and after the maximum (dashed-dotted). The saturation in energy is reached, when the induced charge density reaches its maximum. Beyond the saturation, the induced charge density decreases, resulting in the decrease of average energy.

\subsection{Second order: approximations} \label{sec:chi2}

\begin{figure*}[th]
\center
\begin{minipage}[b]{\columnwidth}
\center
       \begin{minipage}[b]{0.44\columnwidth}
        \includegraphics[width=\columnwidth]{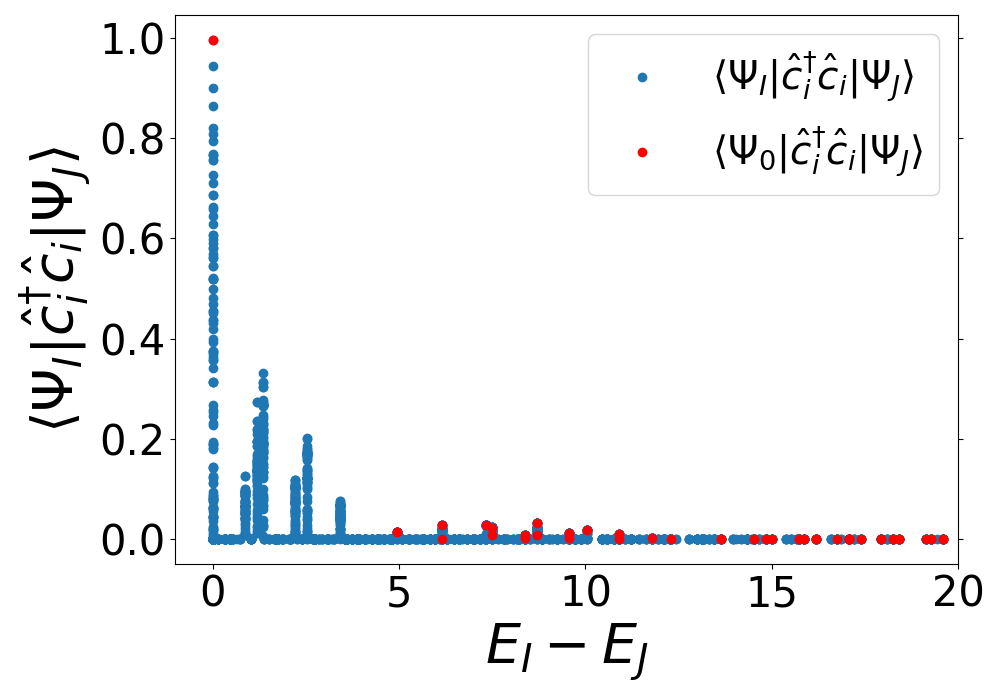}
        \end{minipage}
       \begin{minipage}[b]{0.44\columnwidth}
        \includegraphics[width=\columnwidth]{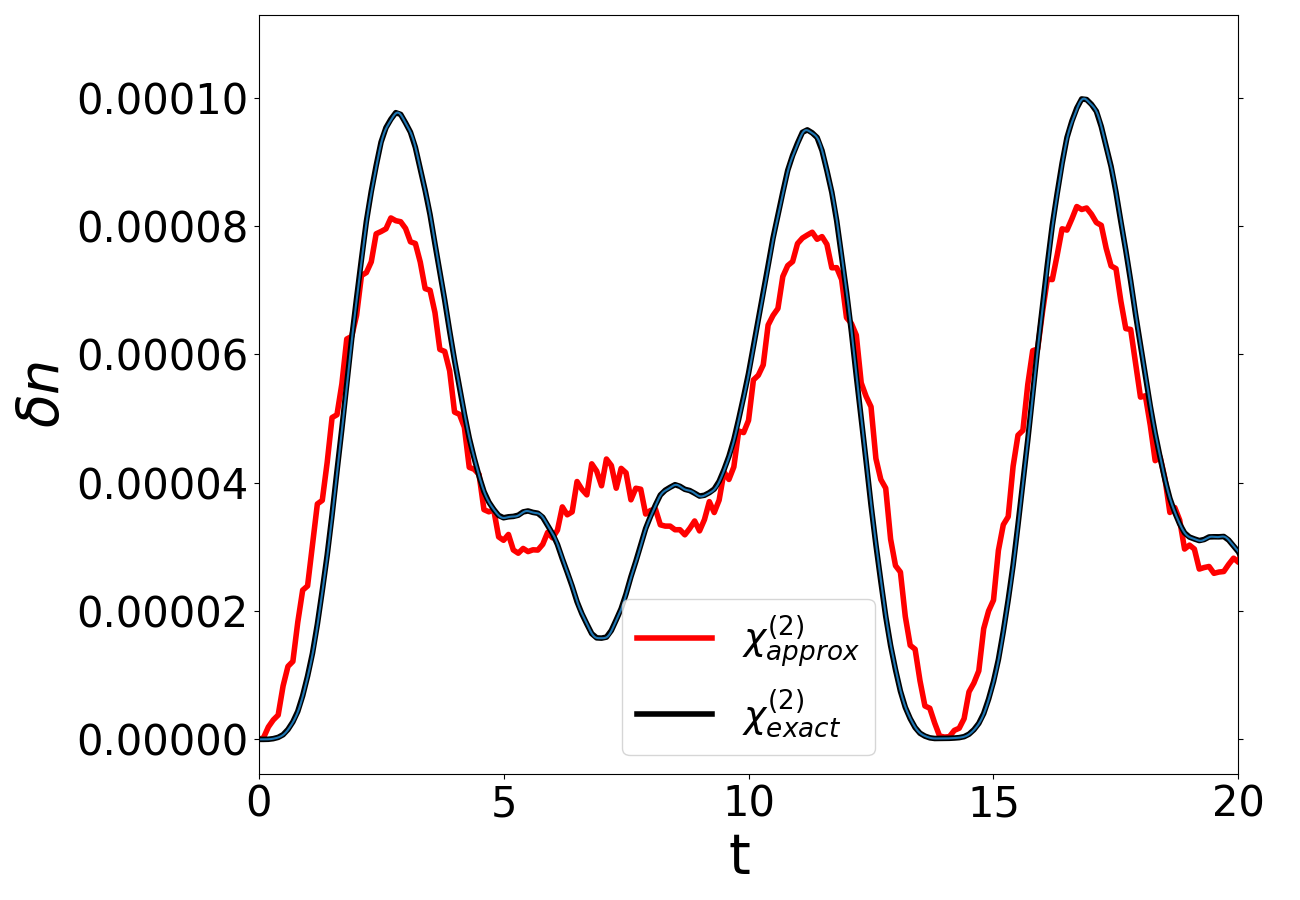}
       \end{minipage}
\end{minipage}
	\caption{\small{
Left: matrix element $\langle \Psi_I | \hat{c}^{\dagger}_i \hat{c}_i | \Psi_J \rangle$ computed on the last electron site of the system of 4 valence and 4 conduction sites (far from the perturbation) as a function of energy difference $E_J-E_I$. In red only the contributions between the ground and excited states are selected.
Right: An approximation to the second order, consisting in keeping only the diagonal elements of the density matrix or, equivalently, only the first line of Eq.~\ref{eq:rho2sin_final}.} 
\label{fig:approx}}
\end{figure*}

Calculating the second-order response function is computationally demanding, so approximations are necessary. Its structure is analyzed here and feasible approximations are recommended. Firstly, spatial and temporal integrals should be handled in advance, either analytically for temporal perturbations (as in this work) or numerically for spatial perturbations when modeling real materials. It will help to avoid the calculation of the entire response function, limiting it to the elements that couple to the external perturbation.

A primary bottleneck in calculating second order response is the summation over excited states (Eq.~\ref{eq:rho2sin_final}). Note that here the matrix elements with the ground state (red points in Fig.~\ref{fig:approx} left) are small compared to the matrix elements between different excited states. % since \vitaly{??? in general the electron density is larger than the transition density \cite{}}. 
The matrix elements with the ground state are present in the linear response Eq.~\ref{eq:chi1obs} and in the last term of the second order Eq.~\ref{eq:rho2sin_final} via $O_{0I}(\rv)$. These linear-response-like terms are then neglected without any effect on the results. This approximation reduces drastically the amount of computation, now one only needs to compute the wings of the spacial integral with the perturbation $\tilde{V}_{0J}$.
The matrix elements entering the second term of Eq.~\ref{eq:rho2sin_final}, \(\langle \Psi_I | \hat{c}^{\dagger}_i \hat{c}_i | \Psi_J \rangle\) tend to vanish for energetically distant states, supporting the introduction of a cutoff (here is at $E_I - E_J \sim 11t$ see Fig. \ref{fig:approx} left). Left panel of Fig.~\ref{fig:approx} also confirms the importance of the first term in Eq.~\ref{eq:rho2sin_final}, since the biggest contribution comes from the diagonal matrix elements. %\vitaly{you can obtain this approximation from the linear response [48,49] of \cite{Dar2025} \cite{}}

An important approximation is obtained by only keeping the first line in Eq.~\ref{eq:rho2sin_final}, which means expanding the time-dependent wave functions on both sides of Eq.~\ref{eq:timepropbasis} only up to the first order.  Only diagonal elements of the density matrix of Eq.~\ref{eq:timepropbasis} are introduced, which should capture the non-coherent processes and result in charge transport and separation. The quality of this approximation can be seen in Fig.~\ref{fig:approx} (right bottom). Indeed, it qualitatively captures the second order response, this has been verified for a large range of perturbation intensities and frequencies. The CPU time for this approximation scales as the linear response one with system size (see Appendix~\ref{app:scaling}). Moreover, as discussed earlier, all the ingredients can be obtained within the linear response TDDFT.

\subsection{Interaction}\label{sec:interaction}

Finally, it is interesting to explore whether the Coulomb interaction changes the validity of linear and quadratic response. The interaction primarily modifies the eigenenergies, lifting some degeneracies, but can also modify the eigenfunctions. Interaction can result in the suppression of the charge dynamics, as it was demonstrated in Ref.~\cite{Gemmer2017} for the Hubbard model. It can also create new spacial pathways for charge dynamics in molecules, which was shown in Ref.~\cite{Despre2022} by \textit{ab-initio} simulations. 

\begin{figure*}[th]
\center
\begin{minipage}[b]{\columnwidth}
\center
       \begin{minipage}[b]{0.45\columnwidth}
	\includegraphics[width=\columnwidth]{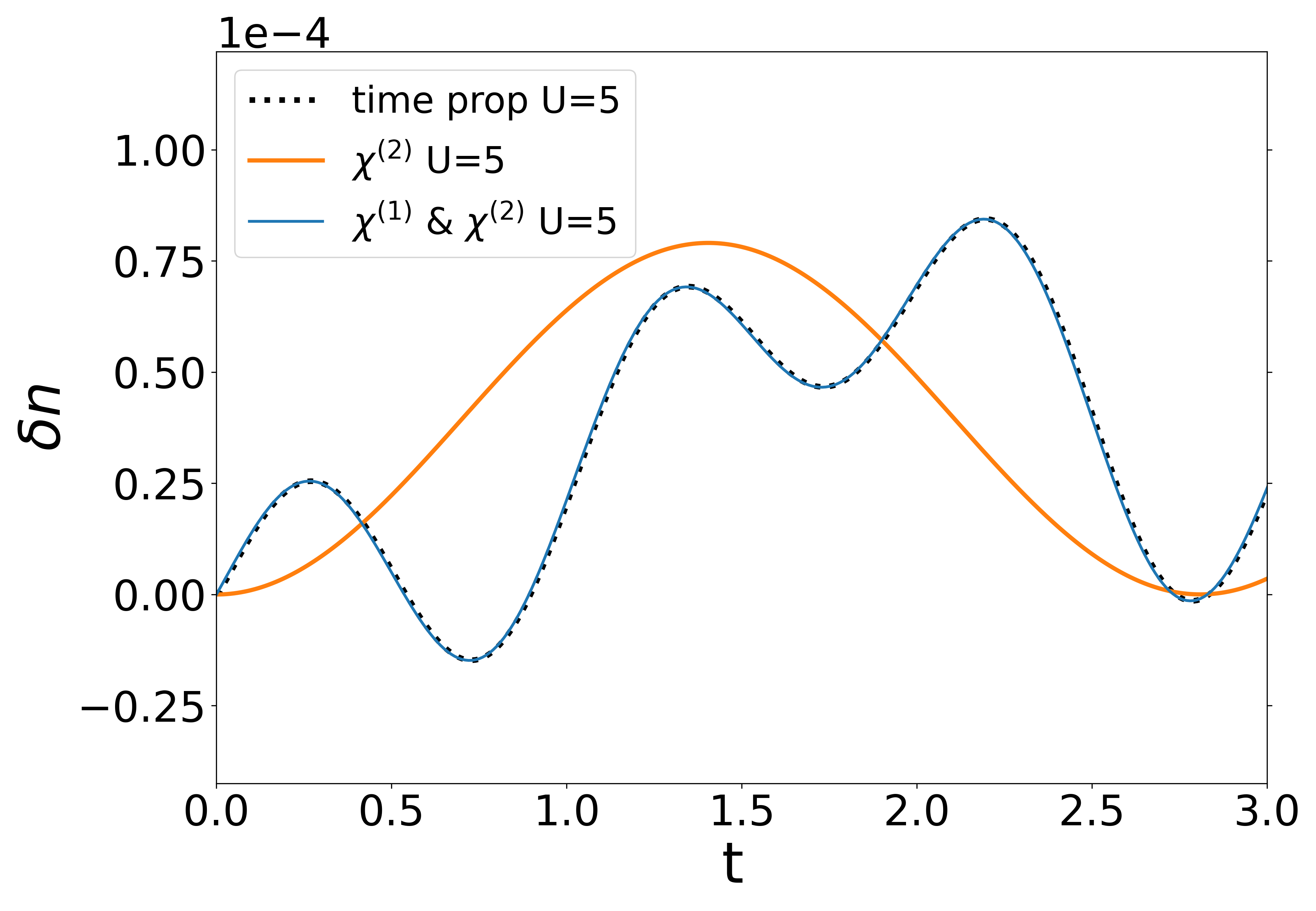}
        \end{minipage}
       \begin{minipage}[b]{0.45\columnwidth}
        \includegraphics[width=\columnwidth]{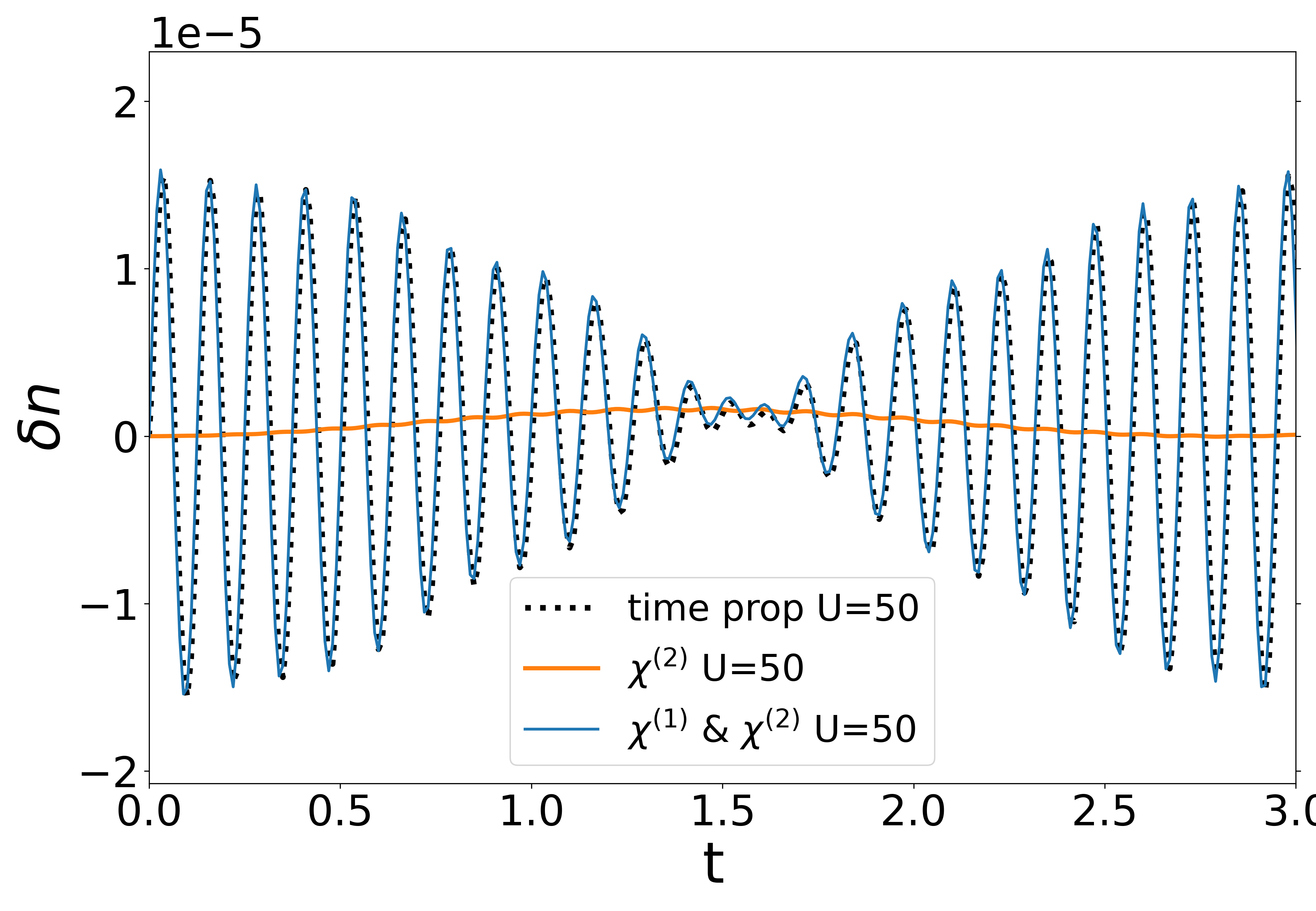}
        \end{minipage}
\end{minipage}
	\caption{\small{Effect of the interaction on charge dynamics. The perturbation is $e(t)= 0.1t_h \cdot \sin(\omega t)$. Left: moderate interaction U=5. Right: Strong interaction U=50} 
\label{fig:interaction}}
\end{figure*}

Here, the effect of onsite interaction $U\sum_i \hat{n}_{i \downarrow} \hat{n}_{i \uparrow}$ for an instantaneous delta-kick perturbation is examined. The spin is introduced, doubling the number of electrons to $N_e=2(N+1)$. The response and the exact time propagation are computed from the many-body states. In the current model, Fig.~\ref{fig:interaction}, which shows the electron layer occupation, one can observe a general reduction of the charge density flow by increasing the interaction, as in Ref.~\cite{Gemmer2017}.

One of the effects of the interaction is altering the ground state, strong interaction prevents the double occupancy of sites, which modifies the validity range of response theory. Based on the overlap interpretation (Fig.~\ref{fig:overlap}), a more extended ground state results in a more accurate linear response. Indeed, in the strong-interaction regime (right panel), where the ground state is delocalized over all sites, linear response dominates. In contrast, for weak interactions, the ground state remains localized on valence sites, enhancing the second-order response in the transport layer.

\section{Conclusions}\label{sec:conclusion}

Time propagation of a wave function for a system under perturbation allows one to access the time evolution of all observables. In this article we have focused on the charge dynamics, charge separation, current and energy transfer under a various weak perturbation, pertinent to optoelectronic and photovoltaic applications.  
We have explored and benchmarked a response theory framework to compute these quantities. Using a numerically solvable site model, we demonstrated that linear response accurately describes coherent charge oscillations under weak perturbations but fails to capture any net charge separation. In contrast, quadratic response theory includes off-diagonal density matrix contributions that enable asymmetric charge propagation and net transport, providing excellent agreement with exact time propagation over a broad range of conditions.

We identified a clear criterion for the validity of second-order response: the amplitude-to-resonance ratio $I/(\Delta_{IJ} \pm \omega) < 1$. This sets the boundary where perturbative approaches remain reliable, with broader validity in non-resonant regimes. We introduced practical approximation to the second order response, that only requires quantities that can be obtained within the linear response TDDFT, which drastically reduces computational demands with minimal loss of accuracy. Second-order response also correctly captures DC components in the current (shift currents), absent in linear response, and remains robust even as density dynamics begin to diverge. Notably, we find that strong on-site interactions extend the range of linear response applicability by delocalizing the ground state.

These results highlight the predictive power of the response theory up to the second order. They also show that with respect to long-time propagation the response theory may be more efficient with respect to full time propagation. While its formal scaling is worse - due to the double sum over states - response functions need only be computed once and allow access to all time scales and perturbations. In practice, symmetries, cutoffs, and proposed controlled approximations significantly reduce computational cost. We expect the response theory to hold for weak perturbations when exploring realistic systems with interfaces and interactions such as electron–hole or electron–phonon coupling. % \vitaly{in which parameter range?}. 

This work highlights the possibility of response theory for accurately simulating  spatially resolved charge dynamics in systems influenced by varying external perturbations, making it a valuable tool for studying fundamental questions and for contributing to the technological advancement in photovoltaic and optoelectronic applications.

\section*{Acknowledgements}
We thank Matteo Gatti and Valérie Veniard for useful discussions. 

% TODO: include author contributions
\paragraph{Author contributions}
L.L., L.R., and V.G. conceived the project and explored the theoretical framework. L.L. and V.G. implemented the simulation code. V.G. carried out the simulations and data analysis. All authors discussed the results and contributed to writing the manuscript.
%This is optional. If desired, contributions should be succinctly described in a single short paragraph, using author initials.

% TODO: include funding information
\paragraph{Funding information}
This work benefited from the support of EDF in the framework of the research and teaching Chair “Sustainable energies” at Ecole Polytechnique (V.G.) and from the European Union’s Horizon 2020 research and innovation programm under the Marie Sklodowska-Curie grant agreement No. 101030447 (L.L.).
%uthors are required to provide funding information, including relevant agencies and grant numbers with linked author's initials. Correctly-provided data will be linked to funders listed in the \href{https://www.crossref.org/services/funder-registry/}{\sf Fundref registry}.

\begin{appendix}
\numberwithin{equation}{section}

	\section{Changing perturbation extension}\label{app:vfull}

\begin{figure*}[th]
\center
\begin{minipage}[b]{\columnwidth}
\center
       \begin{minipage}[b]{0.45\columnwidth}
	       \includegraphics[width=\columnwidth]{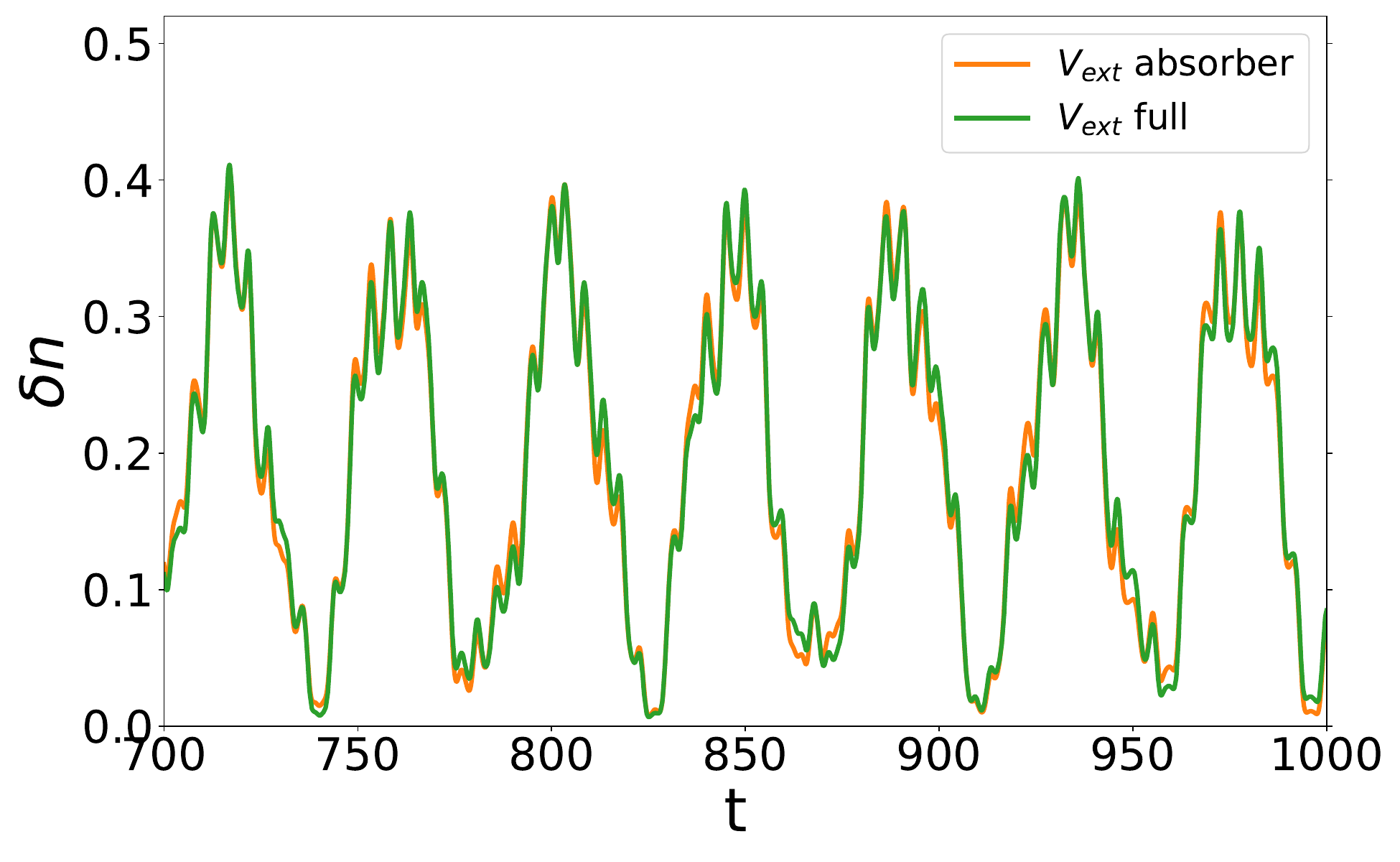}
        \end{minipage}
               \begin{minipage}[b]{0.45\columnwidth}
	       \includegraphics[width=\columnwidth]{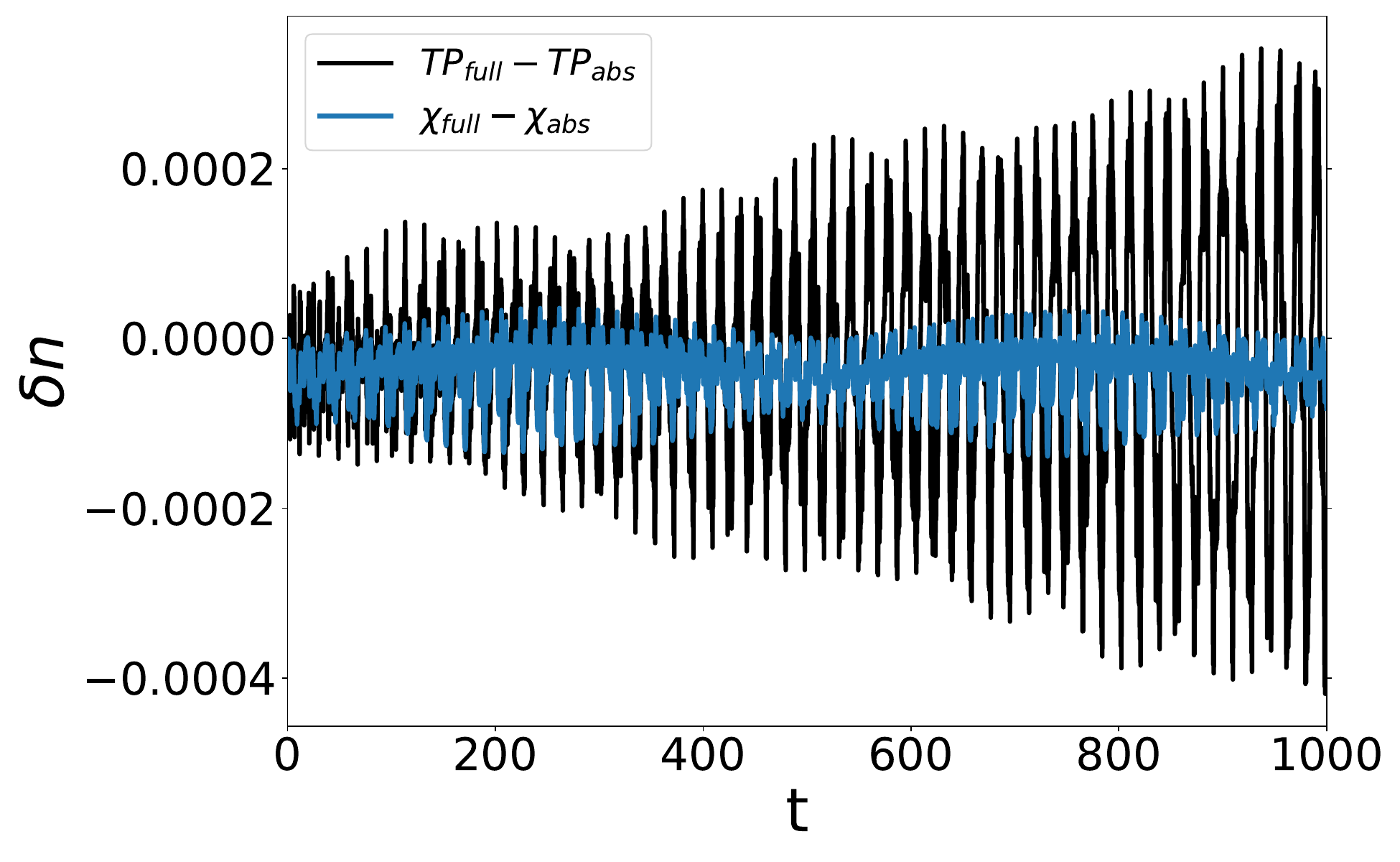}
        \end{minipage}

\end{minipage}
	\caption{\small{The effect of a localised (on absorber) and delocalised perturbation on charge dynamics of the occupation of the last site on the electron transport layer. The perturbation amplitude is 0.3$t_h$, where the third order starts to become important. Left: computed in exact time propagation (TP) under extended perturbation (green) and localized on the absorber (orange). Right: differences between TP under extended and localized perturbations (black) and the same difference in the response theory (blue).} 
\label{fig:Vfull}}
\end{figure*}

	Figure~\ref{fig:Vfull} (left panel) shows the effect of a delocalized external perturbation \\ $\hat{V}_{ext}^{ij}(t)= I \delta_{j i+1} ( \hat{c}^{\dagger}_{i} \hat{c}_{i} + \hat{c}^{\dagger}_{i} \hat{c}_{i+1}+\hat{c}^{\dagger}_{i+1} \hat{c}_{i}) e(t)$ on the dynamics of an occupation at electron layer $c1$ under strong driving ($I = 0.3t_h$), where second order start to break down. For a localized perturbation (orange), only the absorber region is perturbed, as throughout this work. In contrast, a fully delocalized perturbation (green) acts on all sites. The difference arises from additional matrix elements $\tilde{V}_{IJ}$ in the third-order response (see Eq.~\ref{eq:dm3}) and is a purely third-order effect, becoming relevant at longer propagation time. Right panel of Fig.~\ref{fig:Vfull} shows that the response theory (blue line) cannot capture the differences between the delocalized perturbation and localized one obtained in exact time propagation (black line). This means that the delocalized perturbation can shift the regime, where the second order is valid into weaker perturbations. 

	\section{Scaling with system size}\label{app:scaling}

\begin{figure*}[th]
\center
\begin{minipage}[b]{\columnwidth}
\center
       \begin{minipage}[b]{0.45\columnwidth}
	       \includegraphics[width=\columnwidth]{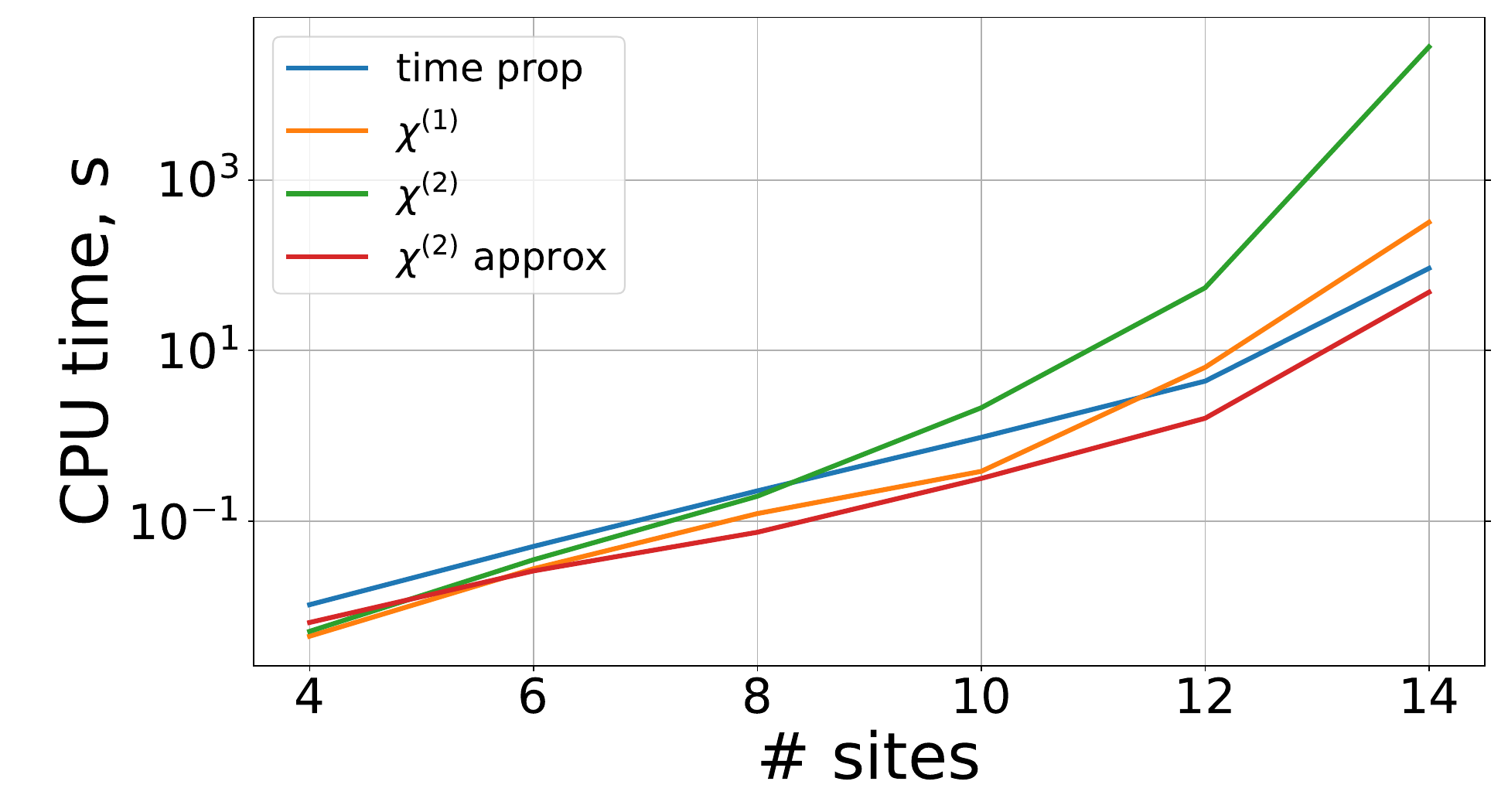}
        \end{minipage}
\end{minipage}
	\caption{\small{CPU time to compute one step of exact dynamics (blue), full scaling of first ($\chi^{(1)}$) (orange), second ($\chi^{(2)}$) (green) order response and an approximation to second order response ($\chi^{(2)}$ approx) (red) as a function of number of sites. Each calculation is performed on 1 CPU core.} 
\label{fig:timescale}}
\end{figure*}

Figure~\ref{fig:timescale} shows CPU time scaling with system size. For exact time propagation (blue), each time step involves a full diagonalization, scaling as $\mathcal{O}(N^3)$ with Hilbert space size $N$. Linear response (orange) requires matrix-vector products and a single sum over states, also scaling as $\mathcal{O}(N^3)$. Quadratic response (green) involves a double sum, leading to $\mathcal{O}(N^4)$ scaling. The approximate second-order scheme from Sec.~\ref{sec:chi2} reduces this to $\mathcal{O}(N^3)$ (red). The the full second-order computation time can be reduced in practice by considering cutoffs in the summations and using symmetries. %\vitaly{One should also note that in practice the second order is also obtained from the second order Dyson equation from the linear response.}
Moreover, the response functions are naturally parallelizable. Instead, time propagation cost is rather a lower limit, since it requires a certain number of time steps and a time-step size convergence study. Importantly, unlike time propagation, once computed, response functions give access to all time scales and allow changes in external perturbations without redoing the full calculation.

%\section{About references}
%Your references should start with the comma-separated author list (initials + last name), the publication title in italics, the journal reference with volume in bold, start page number, publication year in parenthesis, completed by the DOI link (linking must be implemented before publication). If using BiBTeX, please use the style files provided  on \url{https://scipost.org/submissions/author_guidelines}. If you are using our LaTeX template, simply add
%\begin{verbatim}
%\bibliography{your_bibtex_file}
%\end{verbatim}
%at the end of your document. If you are not using our LaTeX template, please still use our bibstyle as
%\begin{verbatim}
%\bibliographystyle{SciPost_bibstyle}
%\end{verbatim}
%in order to simplify the production of your paper.
\end{appendix}


% Provide your bibliography here. You have two options:

%%% FIRST OPTION
% Write your entries here directly, following the example below, including:
% Author(s), Title, Journal Ref. with year in parentheses at the end, followed by the DOI number.

%\begin{thebibliography}{99}
%\bibitem{1931_Bethe_ZP_71} H. A. Bethe, {\it Zur Theorie der Metalle. i. Eigenwerte und Eigenfunktionen der linearen Atomkette}, Zeit. f{\"u}r Phys. {\bf 71}, 205 (1931), \doi{10.1007\%2FBF01341708}.
%\bibitem{arXiv:1108.2700} P. Ginsparg, {\it It was twenty years ago today... }, \url{http://arxiv.org/abs/1108.2700}.
%\end{thebibliography}

% Use your bibtex library, formatted by the SciPost style file.
\bibliography{main.bib}

%%%%%%%%%% END TODO: BIBLIOGRAPHY

\end{document}